\documentclass[prb,twocolumn,showpacs,amsmath,amssymb,floatfix]{revtex4}
\usepackage{graphicx}
\usepackage{graphics}
\usepackage{dcolumn}
\usepackage{bm}
\usepackage{amsmath}
\usepackage{latexsym}
\usepackage{amsfonts}
\usepackage{amssymb}
\usepackage[rightcaption]{sidecap}
\usepackage{color}
\usepackage{float}
\makeatletter

\newcommand{\Rmnum}[1]{\expandafter\@slowromancap\romannumeral  #1@}

\makeatother
\begin{document}
\title{Kadanoff-Baym description of Hubbard clusters out of equilibrium: performance of many-body schemes, correlation-induced damping and multiple steady states.}
\date{\today}
\author{M. Puig von Friesen}
\author{C. Verdozzi}
\author{C.-O. Almbladh}
\affiliation{Mathematical Physics and European Theoretical Spectroscopy Facility (ETSF), Lund University, 22100  Lund, Sweden}
\begin{abstract}
We present in detail a method we recently introduced  
(Phys. Rev. Lett. 103, 176404 (2009)) 
to describe finite systems in and out of equilibrium, 
where the evolution in time is performed via the Kadanoff-Baym 
equations within Many-Body Perturbation theory. The main non-equilibrium 
property we analyze is the time-dependent charge density.
An other quantity we study is the exchange-correlation potential of 
time-dependent Density Functional Theory, obtained via reverse engineering from the time-dependent density.
Our systems consist of small, strongly correlated clusters, 
described by a Hubbard Hamiltonian within the Hartree-Fock, 
second Born, GW and T-matrix approximations. We compare the results from the Kadanoff-Baym dynamics 
to those from exact numerical solutions. The outcome of our comparisons is that, among
the many-body schemes considered, the T-matrix approximation is overall superior at all electron densities.
Such comparisons permit a general assessment of
the whole idea of applying Many-Body Perturbation Theory, in the Kadanoff-Baym sense, to finite systems. 
A striking outcome of our analysis is that when the system evolves under a strong external field, 
the Kadanoff-Baym equations develop a steady-state solution 
as a consequence of a correlation-induced damping. This damping is present both in
isolated (finite) systems, where it is purely artificial, as well as in clusters contacted to (infinite) macroscopic leads. 
To illustrate this point we present selected results for a system coupled to 
contacts within the T-matrix and second Born approximation.
This damping behavior is intrinsically 
linked to the Kadanoff-Baym time-evolution and is not a simple consequence of possible limitations/approximations 
in the calculation of the initial state. 
The extensive numerical characterization we performed indicates that this behavior
is present whenever approximate self energies, based upon infinite partial summations, are used.
A second important result is that, for isolated clusters, the steady state reached is not unique 
but depends on how one switches on the external field. Under some circumstances this is also true for 
clusters connected to macroscopic leads. We provide some statements of more general and conceptual 
character on how the damping mechanism depends 
on the system size and the number of particles, and conclude with an outlook and glimpses of future work.
\end{abstract}
\maketitle
\section{Introduction \label{sec_introduction}}
The Kadanoff-Baym equations (KBE) \cite{KB-book, Keldysh} are one of the fundamental theoretical schemes  
of a microscopic description of quantum systems out of equilibrium.\cite{Danielewicz,Bonitzbook}
Due the growing interest in time-dependent phenomena,
in recent years the KBE have been the object of considerable attention in several
branches of physics. Notable applications of the KBE are in the areas of 
molecular quantum transport, high energy 
coupled  plasmas, nuclear matter, astrophysics, to mention a few.\cite{bonitz1, Jauho, Stefanucci04, Robert, Bonitzdots, Galperin, Thygesen1, Olevano, RobertHubb, Robertlong, Freericks, pva} Another favorable element to the widespread use of KBE is the constantly expanding 
capability of today's computers, which has made the full numerical solution of the KBE possible.
A main strength of the KBE is that one can, in a constructive way, build 
approximations of increasing complexity for the one-particle Green's function, $G$, the key quantity in the KBE. 
These approximations are obtained via Many-Body Perturbation Theory (MBPT) and are
known as conserving approximations, since they guarantee the conservation of important 
quantities such as total energy, number of particles, linear and angular momentum. 

However, the fulfillment of such conserving conditions is no guarantee of the quality of the actual results obtained within a specific
MBA. Hence, it would be useful to have a way to assess the performance of a given conserving MBA.
One of the aims of this paper is to evaluate the range of validity of a group of well known Many-Body Approximations (MBA:s) 
by comparing the one-particle densities
with the exact results for finite strongly correlated clusters out of equilibrium. The main attractive 
feature of such comparisons to exact results is the possibility of 
scrutinizing the performance of the Many-Body Approximations in the non-equilibrium regime. 
This knowledge is most valuable if one wishes to use approximate many-body schemes for systems
(typically, infinite ones, e.g. systems coupled to macroscopic leads as those we consider towards the end of the paper)
where exact solutions are not available.

Describing small clusters with strong electronic correlations, and subject to time-dependent fields, presents interest
not only from a conceptual point of view; there are very many instances
where the technological relevance of cluster physics is manifestly evident.\cite{dots, atomtronics}
The main focus of this work is to investigate the dynamics of finite clusters. 
In this regard, we provide 
here a detailed account of a study performed very recently, \cite{pva} producing additional results
for clusters and presenting in great detail the methodology we developed. We will, however, 
present some results for a system contacted to macroscopic contacts to generalize some of our main findings.
Similar clusters to those discussed here, coupled to leads have already been considered 
either in the stationary limit \cite{Thygesen1} or in the real-time domain. \cite{RobertHubb,Robertlong}

As specific finite model systems, we consider open-ended linear chains \cite{pva1} with short ranged Hubbard interactions.
We study their dynamics through exact diagonalization methods and by propagating the KBE for different MBA:s. 
The approximations we consider are the Hartree-Fock (HFA), 
the second Born (BA), the GW (GWA) \cite{Hedin} and the T-matrix (TMA).\cite{TMA, BonitzTMA}
All these approximations are conserving, \cite{KB-book} which clearly is of great importance when propagating the KBE,
and all of them, apart from HFA, have self energies which are non-local in space and time.
For the GWA, we will consider both a spin-independent and a spin-dependent version. \cite{Thygesen1,Thygesen2} 
The latter has the advantage to alleviate the effect of self-screening. \cite{Pina, Rex}

A general outcome of our study is that the TMA is seen to perform better than the other MBS:s at all fillings and interaction strengths.

With the exact and KBE cluster dynamics at our disposal, we also investigate numerically a well 
established relation between MBPT and another framework to treat 
non-equilibrium phenomena, i.e. time-dependent Density Functional Theory (TDDFT). \cite{Hardy,TDDFT}
Using a spin-independent TDDFT description for the spin-compensated Hubbard model,\cite{VerdozziPRL} 
we will obtain the exchange-correlation (xc) potentials corresponding to the different MBA:s via 
reverse engineering from the time-dependent densities.

Our results show that the
time-dependent KBE present two interesting features. The first is that 
for large external fields, the KBE time evolution in clusters exhibits a damped behavior
induced by many-body correlations (hereafter, we refer to this as correlation-induced damping). 
The second feature is that the steady state one reaches is not unique but depends on how the perturbation is switched on.
We also investigate clusters connected to macroscopic leads, which we in this paper we treat within the TMA and BA.
In this case the presence of the correlation-induced damping is in general not artificial but the steady states may be
non-unique.

These statements are also valid in presence of macroscopic contacts, as shown below for
a small cluster coupled to leads and with a time evolution within the TMA. 

In finite clusters the correlation-induced damping and the existence of multiple steady states is artificial 
and we show that it is due to limitations of self-consistent Many-Body Perturbation theory when applied to finite systems.
We will also provide selected numerical examples, which suggest that correlation-induced damping can be present even in the presence of leads. 

The question whether or not the correlation-induced damping and the existence of multiple steady 
states in contacted systems is a mere consequence of MBPT is at present not so straightforward
to address with full generality.

The paper is organized as follows: we start with a description of our model system(s) in section \ref{Model};
then, in section \ref{sec_Green's function}, we discuss the general properties
of the single-particle Green's function. Section \ref{sec_MBA} is an overview of the Many-Body Approximations used
in this work. Section \ref{sec_groundstate} is devoted to the procedure to obtain the ground state within the 
KBE. How to solve the KBE  for the time evolution is reported in section \ref{sec_time dependence}. In section \ref{sec_TDDFT} we detail how  
we extract the TDDFT exchange-correlation potentials corresponding to a chosen MBA. The ground-state and time-dependent results are 
presented in section \ref{sec_results: groundstate} and \ref{sec_results: time dependence}, respectively. Section \ref{sec_damping} 
deals with the correlation-induced damping which occurs during the KBE time evolution and the existence of multiple steady states. 
Finally, in section \ref{sec_conclusions} we present our conclusions and direction for possible future work.
\section{Model systems} \label{Model}

We will consider one-dimentional clusters with $M$ sites and with one orbital at each site.
Thus, each site (or, equivalently, each orbital) can accommodate a maximum of two electrons, with opposite spin.
The clusters may either be isolated, in which case the Hamiltonian in standard notation is (we set the on-site energies equal to 0)
\begin{equation}
H_{C}\!\!=-V\!\!\!\sum_{\left\langle RR'\right\rangle\!,\, \sigma}\!\!\!\! a_{R\sigma}^{\dagger}a_{R'\sigma}+
U\sum_{R}\!\hat{n}_{R\uparrow}\hat{n}_{R\downarrow}+\!\!\sum_{R,\,\sigma}w_{R}\left(t\right)\hat{n}_{R\sigma},
\label{Hubbard H}\end{equation}
or they may be attached to non-interacting leads of infinite size, as discussed below.
In Eq. (\ref{Hubbard H}), $\hat{n}_{R\sigma}=a_{R\sigma}^{\dagger}a_{R\sigma}$, $\sigma=\uparrow,\downarrow$,
and $\left\langle RR'\right\rangle $ denotes pairs of nearest neighbor sites.
The hopping parameter $V=1$ and $w_R\left(t\right)$ is a local external
field which can be of any shape in time $t$ and space. $U$ and $w_R\left(t\right)$
are given in units of $V$. We will consider clusters with $M=2,4,6$ sites
and, without leads, $N_{e}=2,6$ electrons (in the presence of
leads, the average number of electrons in the clusters is in general non-integer).
Our approach is valid for systems which are compensated as well
as uncompensated in spin. However, in what follows we will only consider clusters (with/out leads)
with an equal average number of spin-up and -down electrons in the ground state; this will hold at all
times during the dynamics, since $H$ has no spin-flip terms. Henceforth, $n=n_\uparrow=n_\downarrow$,
where, $n_\sigma = N_{\sigma}/M$ and $N_\sigma \equiv  \langle \sum_{R \in C} \hat{n}_{R\sigma}\rangle$.
In the presence of leads, $L$, the Hamiltonian is 
\begin{equation}
H=H_{C}+H_{L}+H_{LC}
\end{equation}
where $H_L$ describes non-interacting one-dimentional semi infinite chains,
\begin{equation}
H_{L}=-V_{L}\!\!\!\!\sum_{\langle RR'\rangle\!,\,\sigma \atop R,R'\in L}\!\!\!\!
a_{R\sigma}^{\dagger}a_{R'\sigma} ,
\end{equation}
and $H_{LC}$ describes hopping between the central region, C, and the leads
\begin{equation}
H_{LC}=-V_{LC}\!\!\!\!\!\!\sum_{\langle RR'\rangle\!,\,\sigma \atop R \in L, R' \in C}\!\!\!\!\!\!a_{R\sigma}^{\dagger}a_{R'\sigma} + h.c.
\end{equation}
For isolated clusters, we use a short iterative Lanczos propagation to obtain the exact time evolution. 
A description of our approach for approximate solutions in isolated and contacted clusters is the object of the next five
sections.
\section{The one-particle Green's functions \label{sec_Green's function}}
The one-particle Green's function is a reduced quantity, containing
much less information than the underlying wave function. In general, it describes
a system connected to a bath with which energy and particle can be exchanged. 
The knowledge of the Green's function
gives access to the expectation values of all single-particle operators,
excitation energies and the total energy of the system.
\begin{figure}[t]
\begin{center}
\includegraphics[width=7.7cm, clip=true]{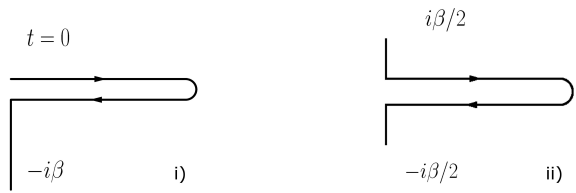}\\
\end{center}
\caption{Keldysh contours.}
\label{paths}
\end{figure}

The general definition of the one particle Green's function is
\begin{eqnarray}
&\hbox{}&\mathcal{G}\left({\bf r}_{1}\sigma_{1}z_{1},{\bf r}_{2}\sigma_{2}z_{2}\right)=
\langle{\bf r}_{1}\sigma_{1}|\mathcal{G}\left(z_{1},z_{2}\right)|{\bf r}_{2}\sigma_{2}\rangle\\
&\hbox{}&\!\!\!\!\!\!-i\left\langle \hat{U}\left(-i\beta,0\right)T_{\gamma}\left[\hat{\psi}_{H}
\left({\bf r}_{1}\sigma_{1}z_{1}\right)\hat{\psi}_{H}^{\dagger}\left({\bf r}_{2}\sigma_{2}z_{2}\right)\right]\right\rangle, 
\label{G definition}\end{eqnarray}
%
%
%
where $\left\langle ...\right\rangle $ denotes expectation values
of the equilibrium ensemble, $\hat{\psi}_{H}$ and $\hat{\psi}_{H}^{\dagger}$
are the field operators in the Heisenberg picture, $\gamma$ is the
Keldysh contour, see Fig. \ref{paths}(i,ii), $T_{\gamma}$ is the path ordering operator, $\hat{U}$
is an evolution operator and $\beta=1/k_{B}T$ is the inverse temperature of the bath.
The $\bf{r}$ and the $s$ denote the labels corresponding to space (site) and 
spin coordinates of the single-particle basis. In this basis the Green's function
becomes a matrix.
The variable $z$ belongs to the Keldysh contour and is in general complex. 
For notational convenience we denote real (imaginary) times by $t$ ($i\tau$).
From the definition of the Green's function one can derive the so-called 
Kubo-Martin-Schwinger condition,\cite{KMS} $\mathcal{G}\left(z_{1},z_{2}\right)=-\mathcal{G}\left(z_{1},z_{2}\pm i\beta\right)$.

The expectation values of all the single-particle operators are obtained according to
\begin{equation}
\left\langle A\left(t\right)\right\rangle =-i\, \textup{Tr}\left[A\left(t\right)\mathcal{G}\left(t,t^{+}\right)\right],\label{observables}
\end{equation}
where the trace, $\textup{Tr} \equiv \sum_{r_1r_2\sigma\sigma'}\delta_{r_1r_2}\delta_{\sigma\sigma'}$, is over space and spin indices. 
As discussed later and in the Appendix A,
the total energy can be found by evaluating 
the Galitskii-Migdal or Luttinger-Ward functionals.

Since in this paper we study only paramagnetic systems in spin-independent fields,
one-particle quantities such as the $G$ and the $\Sigma$ become spin-diagonal:
\begin{equation}
\mathcal{G}\left({\bf r}_{1}\sigma_{1}z_{1},{\bf r}_{2}\sigma_{2}z_{2}\right)=
G\left({\bf r}_{1}z_{1},{\bf r}_{2}z_{2}\right)\delta_{\sigma_{1}\sigma_{2}}. 
\label{G spindiagonal}\end{equation}

In the following, we will use the shorthand notation,
$1=\left({\bf{r}}_{1},z_{1}\right)$, etc., for space time coordinates.

The Green's function obeys an integral equation (the so-called Dyson
equation)
\begin{equation}
G\left(12\right)=G_{0}\left(12\right)+\int_\gamma G_{0}\left(13\right)\Sigma\left(34\right)G\left(42\right)d34,
\label{Dyson time}\end{equation}
with the non-interacting Green's function $G_{0}$ defined by
\begin{equation}
\left(i\partial_{z_{1}}-h\left(1\right)\right)G_{0}\left(12\right)=\delta\left(12\right).\label{Dyson G_0}
\end{equation}
In Eq. (\ref{Dyson G_0}), $h$ is the non-interacting Hamiltonian. \cite{Hartree} It is convenient to decompose $h$ as
$h\left(t\right)=\hat{t}+w\left(t\right)-\mu$, with 
i ) $\hat{t}$ the one-particle kinetic energy ($-\nabla^2_{1}/2$ in coordinate space), 
ii) $w\left(t\right)$ a local external field which  
may depend on time, and iii) $\mu$ the chemical potential. 
The latter is taken to be in between the last occupied and the first non occupied levels. The inclusion
of $\mu$ in $h$ implies that the Fermi energy is placed at zero energy;
electron excitations thus have positive energies, while hole states have
negative. The kernel of the Dyson equation, $\Sigma$, is called the self energy
and is in general non-local in space and time. In the exact theory as well as 
in conserving approximations, \cite{KB-book} the self energy is
a functional of the Green's function, $\Sigma\left[G\right]$, and the
Dyson equation must thus be solved self-consistently. 
In equilibrium all quantities depend only on
$z=z_{2}-z_{1}$ and the equations are then most easily handled, in terms of Fourier transformed quantities,
in the frequency domain.
There, the corresponding Dyson equation becomes a simple matrix equation, involving matrix multiplications; in the single-particle basis,
\begin{equation}
G\left(\epsilon\right)=G_{0}\left(\epsilon\right)+G_{0}\left(\epsilon\right)\Sigma\left(\epsilon\right)G\left(\epsilon\right).
\label{Dyson frequency}\end{equation}
In this paper, we will work only with systems for which the initial state is
the actual ground state (i.e. $\beta\rightarrow\infty$). 
That is, thermal mixtures ($\beta < \infty$)
of initial states will 
not be considered. 
In equilibrium, all two-point propagators can be expressed in terms of a 
spectral function. Specializing to the Green's function, the spectral decomposition has the form

\begin{equation}
G\left(\epsilon\right)=\int\frac{A\left(\epsilon'\right)}{\epsilon'-\epsilon+i\eta\, \textup{sgn}\left(\epsilon'\right)}d\epsilon',
\label{Spectral decomposition}\end{equation}
where the spectral function, $A\left(\epsilon\right)$ is related to the anti-Hermitian part of the corresponding propagator, which for the $G$ is: 
$A\left(\epsilon\right)=-\pi^{-1}\left[G\left(\epsilon\right)-G^{\dagger}\left(\epsilon\right)\right]\textup{sgn}\left(\epsilon\right)$.
The Fermionic spectral function are positive definite and the one for $G$ is normalized:
\begin{equation}
\int A\left(\epsilon\right)d\epsilon=1,
\label{Normalization spectral function}\end{equation}
where 1 represents the identity matrix in the single-particle basis.
%
\section{Many-body Approximations \label{sec_MBA}}

\begin{figure}[h]
\begin{center}
\includegraphics[width=7.7cm, clip=true]{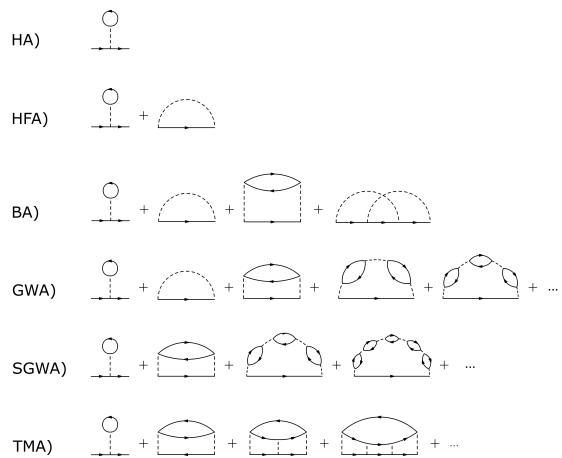}\\
\end{center}
\caption{Diagrammatic expansions of Many-Body Approximations.}
\label{diagrams}
\end{figure}

In general one can not construct the exact self energy and thus needs to rely on approximate schemes.
In Many-Body Perturbation theory (MBPT), one can systematically construct
self energies of increasing complexity. The main idea is making a
diagrammatic expansion of the self energy, and selecting different classes
of diagrams which are then summed up to infinite order. There is a
very important group of approximations which conserve quantities such as the total energy,
the number of particles, linear and angular momentum, when the system is subject to 
external fields. These conserving
properties are related to the fact that the self energy is a functional
derivative of a generating functional $\Phi$,\cite{KB-book}
\begin{equation}
\Sigma\left(12\right)=\frac{\delta\Phi}{\delta G\left(21\right)}.
\label{self energy as functional derivative}\end{equation}
The use of conserving approximations is in general very important
and, in fact, practically mandatory when studying non-equilibrium phenomena.
In this paper we will study the conserving Hartree-Fock,
second Born, GW and T-matrix approximations (HFA, BA, GWA and TMA respectively), see Fig. \ref{diagrams}.
The bare interaction is taken to be local in time, $\mathcal{U}\left(\bf{r}_{1},\bf{r}_{2}\right)\delta\left(t_{1},t_{2}\right)$. 
The formalism we will present is general but we remind the reader that in 
this paper we will only consider local interactions, $\mathcal{U}\left(\bf{r}_{1},\bf{r}_{2}\right)=U\delta\left(\bf{r}_{1},\bf{r}_{2}\right)$
When specialized to our Hubbard clusters with one orbital/site (denoted by $R$), 
the on-site interaction can be treated either as spin-dependent,
$U\,\sum_{R}n_{R\uparrow R\downarrow}$ or as spin-independent 
$\frac{1}{2}U\,\sum_{R\sigma\sigma'}a_{R\sigma}^{\dagger}a_{R\sigma'}^{\dagger}a_{R\sigma'}a_{R\sigma}$.
These two ways are evidently equivalent in any order by order expansion
such as the HFA or the BA. In approximations based upon partial summations, however, this equivalence
may be lost. To illustrate this point we consider the GWA both spin-independently 
(GWA) and spin-dependently (SGWA). The TMA is treated only
spin-dependently.

\textit{Hartree-Fock approximation.} The simplest many-body treatment is given by the Hartree approximation (HA), 
where one takes only the first order direct term into account 
(i.e. the exchange is excluded). Due to its
rather simple nature it will not be consider further. 

The Hartree-Fock approximation (HFA) includes also the first order exchange diagram, the Fock term. The inclusion 
of this diagram, among other things, cures the self interaction of the HA.
The resulting self energy is local in time and thus constant in frequency
space. The remaining diagrams are responsible for the many-body correlations
and give rise to self energies which are non-local in time.

It is often convenient to separate the time-local
HFA and correlation contributions and write
\begin{equation}
\Sigma\left(z_{1},z_{2}\right)=\Sigma_{HF}\:\delta\left(z_{1},z_{2}\right)+\Sigma_{c}\left(z_{1},z_{2}\right).
\label{Split of self energy into local and correlation}\end{equation}

\textit{Second Born approximation.} The simplest scheme which involves correlations is the second Born
approximation (BA) which corresponds to keeping all the diagrams up
to second order. 

\textit{GW approximation.} The GW approximation is the leading term in the expansion of the self energy
in terms of the dynamically screened interaction $W$. The expression
for the self energy in time space is given by
\begin{equation}
\Sigma_{GW}\left(12\right)=\Sigma_{H}+iG\left(12\right)W\left(12\right).
\label{self energy GW}\end{equation}
It should be noted that this expression does not involve any matrix multiplication. The screening of the bare interaction, $U$, results from all possible
electron-hole excitations which are described by a series of bubble
diagrams, involving an irreducible polarization propagator, $P$.
This series can be summed, yielding in frequency space, for a spin-independent interaction,
\begin{equation}
W=\mathcal{U}+\mathcal{U}PW
\label{Dyson W in GWA}\end{equation}
and, for a spin-dependent interaction, 
\begin{equation}
\widehat{W}=\mathcal{U}P\mathcal{U}+\left(\mathcal{U}P\right)^{2}\widehat{W}.
\label{Dyson W in SGWA}\end{equation}
We remind the reader that these Dyson like equations involve matrix multiplications.
In both cases
\cite{Polarization} the polarization propagator, in time space, is 
\begin{equation}
P\left(12\right)=-iG\left(12\right)G\left(21\right).
\label{Polarization}\end{equation}

\textit{T-matrix approximation.} The T-matrix approximation comes from building the T-matrix, $T,$
by summing all the ladder diagrams, representing electron-electron
or hole-hole scattering.\cite{electronhole} The expression for the self energy is given by
\begin{equation}
\Sigma_{TM}\left(12\right)\!=\!\Sigma_{HF}\!+i\!\!\int \mathcal{U}\left(13\right)G\left(43\right)T\left(34\right)\mathcal{U}\left(42\right)d34.
\label{self energy TMA general}\end{equation}
In the case of an on-site, site-independent interaction this simplifies to
\begin{equation}
\Sigma_{TM}\left(12\right)=\Sigma_{HF}+iU^{2}G\left(21\right)T\left(12\right).
\label{self energy TMA}\end{equation}
The sum of the ladder terms in the T-matrix results in
\begin{equation}
T=\phi-\phi \mathcal{U} T,
\label{Dyson T}\end{equation}
where the so-called irreducible vertex $\phi$ is defined as 
\begin{equation}
\phi\left(12\right)=-iG\left(12\right)G\left(12\right).
\label{Vertex}\end{equation}

\section{Ground state \label{sec_groundstate}}

The ground state is obtained by solving the Dyson equation, Eq. (\ref{Dyson frequency})
self-consistently. 
For clusters contacted to non-interacting leads, the problem can be expressed entirely
in terms of propagators which refer only to the central region and an embedding
self energy,\cite{datta,antti}
\begin{equation}
\Sigma_{emb}(\epsilon) = \sum_L |V_{LC}|^2 \tilde{g}_L\left(\epsilon\right) ,
\end{equation}
where $\tilde{g}_L\left(\epsilon\right)$ is the non-interacting Green's function of the uncontacted lead $L$. 
The full Green's function in the central region will now obey a Dyson equation with
both a many-body and an embedding self energy.
The presence of $\Sigma_{emb}$ gives rise to continuous spectra,
and standard techniques can be used to find self-consistent solutions.

For the isolated clusters,
we used a meromorphic representation to be described below.
\subsection{Meromorphic representation of finite systems}
For convenience all one-body quantities are represented as matrices in a 
single-particle basis, e.g. $G_{RR'}\left(\epsilon\right)=\left\langle R\left|G\left(\epsilon\right)\right|R'\right\rangle $. 
In a finite system with a finite phase space, all the spectral functions are discrete:
\begin{equation}
A_{RR'}\left(\epsilon\right)=\sum_{j}A_{RR'}^{j}\delta\left(\epsilon-a_{j}\right),
\label{Meromorph}\end{equation}
where $A_{RR'}^{j}$ is a residue matrix and
$a_{j}$ is a pole position.

>From Eq. (\ref{Spectral decomposition}) we see that the propagators themselves become meromorphic,

\begin{equation}
G_{RR'}\left(\epsilon\right)=\sum_{j}\frac{A_{RR'}^{j}}{\epsilon-a_{j}+i\eta\, \textup{sgn}\left(\epsilon\right)}.
\label{Green's meromorph}\end{equation}

One main advantage in using a meromorphic representation is that
convolutions and cross-correlations are made analytically.\cite{MCCV87} Given the two functions 
\begin{equation}
A_{RR'}\left(\epsilon\right)=\sum_{j}\frac{A_{RR'}^{j}}{\epsilon-a_{j}}\,\,,\,\,B_{RR'}\left(\epsilon\right)=
\sum_{j}\frac{B_{RR'}^{j}}{\epsilon-b_{j}},
\label{A B meromorph}\end{equation}
then their cross-correlation
\begin{equation}
C_{RR'}\left(\epsilon\right)=\int A_{RR'}\left(\epsilon'\right) 
B_{RR'}\left(\epsilon+\epsilon'\right)\frac{d\epsilon'}{2\pi i},
\label{Convolution AB}\end{equation}
becomes
\begin{equation}
C_{RR'}\left(\epsilon\right)=\sum_{j\:k} A_{RR'}^{k}B_{RR'}^{j}\frac{1}{\epsilon+a_{k}-b_{j}}.
\label{Convolution AB analytic}\end{equation}
A second important attractive feature of a meromorphic representation is that one can 
compute at once the equilibrium many-body quantities with any time argument, both real and imaginary. 
For our Hubbard 
clusters, each of the quantities $G,\Sigma,P,W,\Phi,T$ will be expressed in such  
representation during in the actual calculations. 
\subsection{Solution to the Dyson equation}
The solution to the Dyson equation for the $G$ can formally be written
\cite{Dyson}
\begin{equation}
G\left(\epsilon\right)=\left[G_{0}^{-1}\left(\epsilon\right)-\Sigma\left(\epsilon\right)\right]^{-1}=
\left[\epsilon-h-\Sigma\left(\epsilon\right)\right]^{-1}.
\label{Formal solution to Dyson}\end{equation}
The solution of this matrix equation is obtained in two steps \cite{MCCV87}: 1) search
of the pole position and 2) calculation of the residue matrix. \newline
1) The pole positions of $G$ correspond to the zeros of 
\begin{equation}
\det\left[\epsilon-h-\Sigma\left(\epsilon\right)\right],
\label{Determinant G}\end{equation}
which we find by ordinary root finding algorithms.\newline
2) Once the pole positions are found we calculate the residue matrices
by integration in the complex plain. We have in general
\begin{equation}
\oint f\left(\epsilon\right)=2\pi i\sum_{j}\textup{Res}\left(f,a_{j}\right),
\label{Residues theorem}\end{equation}
where 
\begin{equation}
f\left(\epsilon\right)=\sum\frac{A_{j}}{\epsilon-a_{j}}.
\label{Meromorph f}\end{equation}
If we now perform a closed integration around the pole $a_{j}$ we
obtain directly the residue matrix
\begin{equation}
A_{j}=\frac{1}{2\pi i}\oint_{a_{j}}f\left(\epsilon\right).
\label{Resulting residue matrix}\end{equation}
This integration is, in practice, performed numerically.

\subsection{Self-consistency}

To reach self-consistency, we start by constructing the self energy
with some initial $G$, normally taken to be $G_{0},$ and then solve the
corresponding Dyson equation. The resulting $G$ is then used to build
the new self energy and the procedure is carried on until convergence. 
In order to keep the number of poles under control (such number increases rapidly from
iteration to iteration),  we make use of a "decimation" procedure,
where poles are merged if there are small or close enough. When two
poles are merged, the new pole position is the old center of mass
position (``mass'' being the trace of the residue matrix) and
the new residue matrix is the sum of the two old residue matrices. 
To improve the convergence we update $G$ by making a linear combination
of the new (solution of the Dyson equation) and the old $G$:s.
There are different functionals which yield the total energy of a system, 
all of which are equivalent at the point of stationary solution to
the Dyson equation. However, away from self-consistency, they do in general
not coincide. One independent way of evaluating the degree of self-consistency
is thus by comparing the values of different energy functionals, see appendix A.
\section{Time dependence \label{sec_time dependence}}
When an external field is applied to a system, it is in general driven out of
equilibrium. When the system is out of equilibrium and away from the
steady-state regime, all the quantities will intrinsically depend on
the two time arguments $\left(t_{1},t_{2}\right)$ separately and
the Keldysh formalism becomes essential. To explicitly show
which convention we used in this paper for the path-ordered two-point Keldysh functions, $\mathcal{K}$,
we give some definitions. \newline
A general Keldysh function can be written

\begin{eqnarray}
&\hbox{}&\mathcal{K}\left(12\right)=\mathcal{K}^{\delta}\left(12\right)\delta\left(z_{1},z_{2}\right)\nonumber\\
&\hbox{}&\quad\quad\quad+\Theta\left(12\right)\mathcal{K}^{>}\left(12\right)+\Theta\left(21\right)\mathcal{K}^{<}\left(12\right),
\label{Keldysh components}\end{eqnarray}
where $>$ ($<$) refers to the hole (electron) part,
and $\mathcal{K}^\delta$ the time-local part.

The time coordinates
are on the Keldysh contour and may be real or complex.
For real times we may also introduce retarded and advanced propagators,

\begin{eqnarray}
&\hbox{}&\mathcal{K}^{R}\left(12\right)=\mathcal{K}^{\delta}\left(12\right)\delta\left(z_{1},z_{2}\right)\nonumber\\
&\hbox{}&\qquad\quad\quad+\Theta\left(t_{1},t_2\right)\left[\mathcal{K}^{>}\left(12\right)-\mathcal{K}^{<}\left(12\right)\right],\label{Retarded}\\
&\hbox{}&\mathcal{K}^{A}\left(12\right)=\mathcal{K}^{\delta}\left(12\right)\delta\left(z_{1},z_{2}\right)\nonumber\\
&\hbox{}&\qquad\quad\quad-\Theta\left(t_{2},t_{1}\right)\left[\mathcal{K}^{>}\left(12\right)-\mathcal{K}^{<}\left(12\right)\right].
\label{Advanced}\end{eqnarray}
Note that for the $\mathcal{K}$:s considered in the paper, only $\Sigma$ and $W$
have parts which are local in time.

When both time arguments are imaginary, the Keldysh function reduces to the corresponding equilibrium Matsubara function:
\begin{equation}
\mathcal{K}^{M}\left(\tau-\tau'\right)=-i\mathcal{K}\left(-i\tau,-i\tau'\right).
\label{Matsubara}\end{equation} 
In Eq. (\ref{Keldysh components}),
$\Theta\left(12\right)$ should be understood as $\Theta\left(z_{1},z_{2}\right)$,
a generalized Heaviside function for $z_{1}, z_{2}$ on the ordered Keldysh contour.
In is worth noting that when both time arguments lie on the imaginary (Matsubara) axis, 
the quantities represent the initial, equilibrium, state which depend only on the time differences. 
As an example of how these terms are found in equilibrium in the meromorphic representation 
we display the hole contribution to the Green's function:
\begin{equation}
G^{<}\left(t_1,t_2\right)=i\sum_{j<\mu}A_{j}e^{ia_{j}\left(t_1-t_2\right)}
\label{Equilibrium: Both arguments real}\end{equation}
when both time arguments are real,
\begin{equation}
G^{<}\left(t,-i\tau\right)=i\sum_{j<\mu}A_{j}e^{ia_{j}t}e^{a_{j}\tau}
\label{Equilibrium: Mixed arguments}\end{equation}
when both one time argument is real and one imaginary,
\begin{equation}
G^{<}\left(-i\tau_{1},-i\tau_{2}\right)=i\sum_{j<\mu}A_{j}e^{-a_{j}\left(\tau_{1}-\tau_{2}\right)}
\label{Equilibrium: Both arguments imaginary}\end{equation}
when both time arguments are imaginary.
\subsection{General symmetries}
We recall some prominent symmetry relations which will be
used during the time propagation.
>From the definition of the Green's function, Eq. (\ref{G definition}),
one can derive a very important symmetry,\cite{Symmetry} which
enters many relevant and useful relations:
\begin{equation}
G^{\gtrless}\left(12\right)=-G^{\gtrless}\left(21\right)^{\dagger}.
\label{General symmetry G}\end{equation}
Additionally, from the definition of the retarded and advanced
Green's functions we obtain
\begin{equation}
G^{R/A}\left(12\right)=G^{A/R}\left(21\right)^{\dagger}.
\label{General symmetry R/A}\end{equation}
From the expansion of the $T$ and $W$ in terms of $\phi$ and $P$
it follows that the $T$ and $W$ will have the same symmetry properties
as $\phi$ and $P$. The symmetries of $\phi$ and $P$ can be deduced
from their definitions
\begin{equation}
P^{\gtrless}\left(12\right)=-P^{\gtrless}\left(21\right)^{\dagger}\Longrightarrow 
W^{\gtrless}\left(12\right)=-W^{\gtrless}\left(21\right)^{\dagger}
\label{General symmetry W}\end{equation}
and
\begin{equation}
\phi^{\gtrless}\left(12\right)=-\phi^{\gtrless}\left(21\right)^{\dagger}\Longrightarrow 
T^{\gtrless}\left(12\right)=-T^{\gtrless}\left(21\right)^{\dagger}.
\label{General symmetry T}\end{equation}
In a similar way we find a symmetry relation, valid for all approximations, for the self energy,

\begin{equation}
\Sigma^{\gtrless}\left(12\right)=-\Sigma^{\gtrless}\left(21\right)^{\dagger}.
\label{General symmetry self energy}\end{equation}

An additional symmetry fulfilled by $W$ is
\begin{equation}
W^{\gtrless}\left(12\right)=W^{\lessgtr}\left(21\right),\label{Specific symmetry W}\end{equation}
which implies
\begin{equation}
W^{>}\left(z,z\right)=W^{<}\left(z,z\right)\; ,\; \textup{Re}\,W_{RR'}\left(z,z\right)=0\quad\forall\; R,R'.
\label{Relation W on time diagonal}\end{equation}
No equivalent relation exists for $T$.

\subsection{Solving for the non-equilibrium Green's function}

To obtain the non-equilibrium Green's function we need to solve the
corresponding equations of motion, called the Kadanoff-Baym equations
(KBE),
\begin{eqnarray}
&\hbox{}&\!\!\!\!\!\!\!\!\!\!\!\!\!\!\!\!\left(i\partial_{t_{1}}-h\left(1\right)\right)G\left(12\right)=\delta\left(12\right)
+\int_{\gamma}\Sigma\left(13\right)G\left(32\right)d3,\label{KBE left}\\
&\hbox{}&\!\!\!\!\!\!\!\!\!\!\!\!\!\!\!\!\left(-i\partial_{t_{2}}-h\left(2\right)\right)G\left(12\right)=\delta\left(12\right)
+\int_{\gamma}G\left(13\right)\Sigma\left(32\right)d3.
\label{KBE right}\end{eqnarray}

The kernel of these equations, the $\Sigma$, will in general be a
contraction of the Green's function with an other quantity which involves
an infinite order summation such as the $T$ or $W$. These quantities
are defined by corresponding integral equations. Specializing to the case
of  $T$, we have 
\begin{eqnarray}
T\left(12\right)=\Phi\left(12\right)\!-\!\!\int_{\gamma}\!\!\Phi\left(13\right)\mathcal{U}\left(34\right)T\left(42\right)d34,\label{Dyson T time left}\\
T\left(12\right)=\Phi\left(12\right)\!-\!\!\int_{\gamma}\!\!T\left(13\right)\mathcal{U}\left(34\right)\Phi\left(42\right)d34.
\label{Dyson T time right}\end{eqnarray}
We thus have two sets of coupled integral equations which need to
be solved simultaneously at all times.

\subsection{Solution of coupled integral equations}

\begin{figure}[t]
\begin{center}
\includegraphics[width=7.7cm,clip=true]{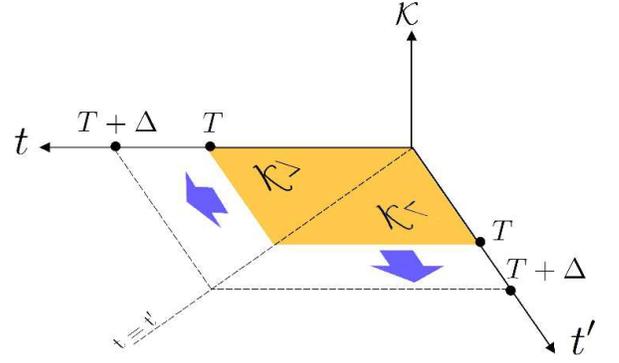}\\
\end{center}
\caption{(Color online) Time square.}
\label{timesquare}
\end{figure}

From the symmetry relations Eqs. (\ref{General symmetry G}, \ref{General symmetry W}, \ref{General symmetry T})
we see that $G$ and $W,\:T$ are only needed on the upper/lower
time matrix. We choose the lesser components on the upper triangle $t_{1}\geq t_{2}$
and the greater ones  on the lower triangle $t_{1}<t_{2}$.\cite{Tsymmetry} 
Propagation is thus made by expanding the Keldysh functions on the time square from $T$
to $T+\Delta$, see Fig. \ref{timesquare}. In the integral equations, 
Eqs. (\ref{KBE left}, \ref{KBE right}, \ref{Dyson T time left}, \ref{Dyson T time right}),
the function to be determined appears in 
both the right and left-hand sides. To solve these equations we use
a self-consistent predictor-corrector method.\cite{Kohler} The method can be described
schematically by an external loop, where an approximate $\bar{G}$
at $T+\Delta$ is generated by propagating Eqs. (\ref{KBE left}, \ref{KBE right}),
and an internal loop, where Eqs. (\ref{Dyson T time left}, \ref{Dyson T time right})
are solved self consistently for a fixed kernel $\phi\left[\bar{G}\right]$.
The external loop is performed by a predictor-corrector method described
below while the internal one is solved by the same iteration method
as described for the ground state. The external loop is initiated
by an extrapolated value of the collision integrals $\int_{\gamma}\Sigma_c G$,
whilst the internal loop is started by an extrapolated value of the
$W,\:T$. The new time step is generated when the external loop achieves
self-consistency.

\subsection{Kadanoff-Baym equations}
\label{sec:KBE}
When propagating the Green's functions it is convenient to separate the terms which are
local in time and single-particle like ($h$ and $\Sigma_{HF}$) from the remaining 
correlation-induced ones\cite{Kohler} and introduce
\begin{equation}
\mathfrak{h}=h+\Sigma_{HF}.
\end{equation}
In this way, $\mathfrak{h}$ subtitutes $h$ and the full self energy is replaced by its correlation part $\Sigma_c$
in the KBE (\textit{cf.} Eqs. (\ref{KBE left}, \ref{KBE right})).
The reason for this partitioning is twofold: on the one hand the contribution 
from the single
particle evolution is very important (and could thus lead to large
numerical errors in the correlation contribution) and on the other
it can be solved essentially in an exact way as will be detailed in Sec \ref{sec:timeprop}.

There are several equivalent contours on which one can define the KBE. 
We use the contour ii) in Fig. \ref{paths}, which is numerically more stable and has an
analytical limit when the temperature goes to zero.
Once we specialize to this contour, the KBE become
\begin{eqnarray}
&\hbox{}&\!\!\!\!\!\!i\partial_{t_{1}}G^\lessgtr\left(t_{1},t_{2}\right)=\mathfrak{h}\left(t_{1}\right)G^\lessgtr\left(t_{1},t_{2}\right)+I_{1}^\lessgtr\left(t_{1},t_{2}\right),\label{KBE greater real}\\
&\hbox{}&\!\!\!\!\!\!-i\partial_{t_{2}}G^\lessgtr\left(t_{1},t_{2}\right)=G^\lessgtr\left(t_{1},t_{2}\right)\mathfrak{h}\left(t_{2}\right)+I_{2}^\lessgtr\left(t_{1},t_{2}\right),\label{KBE lesser real}\\
&\hbox{}&\!\!\!\!\!\!i\partial_{t}G^{<}\left(t,-i\tau\right)=\mathfrak{h}\left(t\right)G^{<}\left(t,-i\tau\right)+I^{<}\left(t,-i\tau\right),\label{KBE lesser imaginary}\\
&\hbox{}&\!\!\!\!\!\!-i\partial_{t}G^{>}\left(-i\tau,t\right)=G^{>}\left(-i\tau,t\right)\mathfrak{h}\left(t\right)+I^{>}\left(-i\tau,t\right),\label{KBE greater imaginary}
\end{eqnarray}
where the collision integrals with both time arguments real are
\begin{eqnarray}
\hbox{}I_{1}^\lessgtr\!\left(t_{1},t_{2}\right)\!=\!\!\!\int_{0}^{t_{1}}\!\!\!d\overline{t}\left[\Sigma_{c}^{R}\!\left(t_{1},\overline{t}\right)G^\lessgtr\!\left(\overline{t},t_{2}\right)+\Sigma_{c}^\lessgtr\!\left(t_{1},\overline{t}\right)G^{A}\!\left(\overline{t},t_{2}\right)\right]\nonumber\\
\hbox{}\!\!\!\!\!\!\!\!\!\!\!\!\!\!\!+\!\frac{1}{i}\!\int_{0}^{\beta/2}\!\!\!\!\!\!\!\!\!d\overline{\tau}\left[\Sigma_{c}^\lessgtr\!\left(t_{1},-i\overline{\tau}\right)G^\lessgtr\!\left(-i\overline{\tau},t_{2}\right)+\Sigma_{c}^\lessgtr\!\left(t_{1},i\overline{\tau}\right)G^\lessgtr\!\left(i\overline{\tau},t_{2}\right)\right],
\nonumber\\
\label{Collision integral greater real}\end{eqnarray}
\begin{eqnarray}
\hbox{}I_{2}^\lessgtr\!\left(t_{1},t_{2}\right)\!=\!\!\!\int_{0}^{t_{2}}\!\!\!d\overline{t}\left[G^{R}\left(t_{1},\overline{t}\right)\Sigma_{c}^\lessgtr\!\left(\overline{t},t_{2}\right)+G^\lessgtr\!\left(t_{1},\overline{t}\right)\Sigma_{c}^{A}\left(\overline{t},t_{2}\right)\right]\nonumber\\
\hbox{}\!\!\!\!\!\!\!\!\!\!\!\!\!\!\!+\!\frac{1}{i}\!\int_{0}^{\beta/2}\!\!\!\!\!\!\!\!\!d\overline{\tau}\left[G^\lessgtr\!\left(t_{1},-i\overline{\tau}\right)\Sigma_{c}^\lessgtr\!\left(-i\overline{\tau},t_{2}\right)+G^\lessgtr\!\left(t_{1},i\overline{\tau}\right)\Sigma_{c}^\lessgtr\!\left(i\overline{\tau},t_{2}\right)\right].
\nonumber\\
\label{Collision integral lesser real}\end{eqnarray}
The collision integrals with one of the time arguments complex specialize to
\begin{eqnarray}
&\hbox{}&I^{<}\left(t,-i\tau\right)=\!\!\int_{0}^{t}\!\!d\overline{t}\Sigma_{c}^{R}\left(t,\overline{t}\right)G^{<}\left(\overline{t},-i\tau\right)\nonumber\\
&\hbox{}&\!\!\!\!+\!\int_{0}^{\beta/2}\!\!\!\!\!\!\!\!\!d\overline{\tau}\left[\Sigma_{c}^{<}\!\left(t,-i\overline{\tau}\right)G^{M}\!\left(\overline{\tau}-\tau\right)+\Sigma_{c}^{>}\!\left(t,i\overline{\tau}\right)G^{M}\!\left(-(\overline{\tau}+\tau)\right)\right],
\nonumber\\
\label{Collision integral lesser imaginary}\end{eqnarray}
\begin{eqnarray}
&\hbox{}&I^{>}\left(-i\tau,t\right)=\!\!\int_{0}^{t}\!\!d\overline{t}G^{>}\left(-i\tau,\overline{t}\right)\Sigma_{c}^{A}\left(\overline{t},t\right)\nonumber\\
&\hbox{}&+\!\int_{0}^{\beta/2}\!\!\!\!\!\!\!\!\!d\overline{\tau}\left[G^{M}\!\left(\tau-\overline{\tau}\right)\Sigma_{c}^{>}\!\left(-i\overline{\tau},t\right)+G^{M}\!\left(\tau+\overline{\tau}\right)\Sigma_{c}^{<}\!\left(i\overline{\tau},t\right)\right].
\nonumber\\
\label{Collision integral greater imaginary}
\end{eqnarray}
It is worth noting that
all Eqs. (\ref{Collision integral greater real} - \ref{Collision integral lesser imaginary}) 
contain terms which involve integration along the Matsubara (vertical) axis 
and which represent the memory of initial state correlations during the time evolution. 

For the collision integrals, one can derive a similar symmetry property:  
\begin{equation}
I_{1}^{\lessgtr}\left(12\right)=-I_{2}^{\lessgtr}\left(21\right)^{\dagger}.
\label{General symmetry collision integral}\end{equation}
An important consequence of this relation is that the densities
\cite{Realdensity} are manifestly real. From the KBE one can derive that the condition
for real densities is given by
\begin{equation}
\left[I_{1}^{<}\left(t,t\right)-I_{2}^{<}\left(t,t\right)\right]=
\left[I_{1}^{<}\left(t,t\right)-I_{2}^{<}\left(t,t\right)\right]^{\dagger},
\label{Condition real density}\end{equation}
which is manifestly satisfied by Eq. (\ref{General symmetry collision integral}).
Furthermore, on the time diagonal, we obtain another relation, which is very useful
in from the computationally point of view: 
\begin{equation}
I_{1/2}^\lessgtr=I_{1/2}^{\gtrless}.
\label{Relation collision integrals on the time diagonal}\end{equation}
From the properties of integral equations, it follows that all the
symmetry and structure properties of the Green's function are those of $G_{0}$.

The discussion of the KBE above is valid for extended and finite systems
alike. When the interaction is confined to a central region which is contacted to
possibly macroscopic leads, the problem can again be expressed in propagators
which refer only to the central region, and an embedding self energy which 
now depends on time,\cite{RobertHubb,Robertlong}
\begin{equation}
\Sigma_{emb}(t_1, t_2) = \sum_L |V_{LC}|^2 \tilde{g}_L\left(t_1, t_2\right) .
\end{equation}
Here $\tilde{g}_L\left(t_1, t_2 \right)$ is the non-interacting Green's function 
of the uncontacted lead $L$, possibly subject to a uniform but time-dependent bias.
The full Green's function in the central region will now obey the KBE  
with both the self energy from the interaction and the one from the leads via the embedding.
The embedding self energy is entirely non-local in time and will be treated on the
same footing as the correlation part of the interaction self energy. It is
worth noting that the embedding self energy involve no self-consistency and
can thus be calculated once the external bias (if present) is known.
\subsection{The Dyson equation for T and W}

Propagating in time the KBE within a specific MBA-based partial summation,
requires solving a Dyson equation for auxiliary quantities which enter the
expression for the self energy. For the TMA and GWA such quantities are the
$T$ and $W$, respectively. The components of the corresponding
Dyson equations (again specialized to the case of $T$) Eqs. (\ref{Dyson T time left}, \ref{Dyson T time right})
are, for both times on the real axis, 
\begin{eqnarray}
&\hbox{}&T^\lessgtr\left(t_{1},t_{2}\right)=\Phi^\lessgtr\left(t_{1},t_{2}\right)\nonumber\\
&\hbox{}&-\!\!\int_{0}^{t_{1}}\!\!d\bar{t}\left[\Phi^{R}\!\left(t_{1},\bar{t}\right)\mathcal{U}T^\lessgtr\!\left(\bar{t},t_{2}\right)+\Phi^\lessgtr\!\left(t_{1},\bar{t}\right)\mathcal{U}T^{A}\!\left(\bar{t},t_{2}\right)\right]\nonumber\\
&\hbox{}&\!\!\!\!\!\!\!\!\!\!-\frac{1}{i}\!\!\int_{0}^{\beta/2}\!\!\!\!\!\!\!\!\!d\bar{\tau}\!\!\left[\Phi^{<}\!\left(t_{1},\!-i\bar{\tau}\right)\mathcal{U}T^{>}\!\left(\!-i\bar{\tau},t_{2}\right)\!+\!\Phi^{>}\!\left(t_{1},i\bar{\tau}\right)\mathcal{U}T^{<}\!\left(i\bar{\tau},t_{2}\right)\right],
\nonumber\\
\label{T greater real}\end{eqnarray}
\begin{eqnarray}
&\hbox{}&T^\lessgtr\left(t_{1},t_{2}\right)=\Phi^\lessgtr\left(t_{1},t_{2}\right)\nonumber\\
&\hbox{}&-\!\!\int_{0}^{t_{2}}\!\!d\bar{t}\left[T^{R}\!\left(t_{1},\bar{t}\right)\mathcal{U}\Phi^\lessgtr\!\left(\bar{t},t_{2}\right)+T^\lessgtr\!\left(t_{1},\bar{t}\right)\mathcal{U}\Phi^{A}\!\left(\bar{t},t_{2}\right)\right]\nonumber\\
&\hbox{}&\!\!\!\!\!\!\!\!\!\!-\frac{1}{i}\!\!\int_{0}^{\beta/2}\!\!\!\!\!\!\!\!\!d\bar{\tau}\!\!\left[T^{<}\!\left(t_{1},\!-i\bar{\tau}\right)\mathcal{U}\Phi^{>}\!\left(\!-i\bar{\tau},t_{2}\right)\!+\!T^{>}\!\left(t_{1},i\bar{\tau}\right)\mathcal{U}\Phi^{<}\!\left(i\bar{\tau},t_{2}\right)\right],\nonumber\\
\label{T lesser real}\end{eqnarray}
when one of the time arguments is imaginary we have
\begin{eqnarray}
&\hbox{}&T^{<}\left(t,-i\tau\right)=\Phi^{<}\left(t,-i\tau\right)-\!\!\!\int_{0}^{t}d\bar{t}\Phi^{R}\!\left(t,\bar{t}\right)\mathcal{U}T^{<}\!\left(\bar{t},-i\tau\right)\nonumber\\
&\hbox{}&\!\!\!\!\!\!\!\!\!\!-\!\!\int_{0}^{\beta/2}\!\!\!\!\!\!\!\!\!d\bar{\tau}\!\!\left[\Phi^{<}\!\left(t,\!-i\bar{\tau}\right)\mathcal{U}T^{M}\!\left(\bar{\tau}-\tau\right)+\Phi^{>}\!\left(t,i\bar{\tau}\right)\mathcal{U}T^{M}\!\left(-\left(\bar{\tau}+\tau\right)\right)\right],\nonumber\\
\label{T lesser imaginary}\end{eqnarray}
\begin{eqnarray}
&\hbox{}&T^{>}\left(-i\tau,t\right)=\Phi^{>}\left(-i\tau,t\right)-\!\!\!\int_{0}^{t}d\bar{t}T^{>}\!\left(-i\tau,\bar{t}\right)\mathcal{U}\Phi^{A}\!\left(\bar{t},t\right)\nonumber\\
&\hbox{}&-\!\!\int_{0}^{\beta/2}\!\!\!\!\!\!\!\!\!d\bar{\tau}\left[T^{M}\!\left(\tau-\bar{\tau}\right)\mathcal{U}\Phi^{>}\!\left(\!-i\bar{\tau},t\right)\!+\!T^{M}\!\left(\tau+\bar{\tau}\right)\mathcal{U}\Phi^{<}\!\left(i\bar{\tau},t\right)\right].\nonumber\\
\label{T greater imaginary}\end{eqnarray}

\subsection{Time propagation algorithm}
\label{sec:timeprop}
As mentioned in section \ref{sec:KBE} we treat the time-local part
of the self energy ($\Sigma_{HF}$)
on the same footing as the non-interacting terms ($h$) in the time propagation.
The evolution from these single-particle terms can be expressed in terms
of a single-particle evolution operator $S$ which is a time-dependent
matrix in single-particle labels. This leads to the following unitary
gauge transformation
\begin{equation}
G^{\lessgtr}\left(t_{1},t_{2}\right)=S\left(t_{1},0\right)g^{\lessgtr}
\left(t_{1},t_{2}\right)S^{\dagger}\left(t_{2},0\right),
\label{Gauge transformation}\end{equation}
where $S$ satisfies
the following differential equation:
\begin{equation}
i\partial_{t_{1}}S\left(t_{1},0\right)=\mathfrak{h}\left(t_{1}\right)S\left(t_{1},0\right),
\label{Equation of motion of sp evolution operator}\end{equation}
with the initial condition
\begin{equation}
S\left(0,0\right)=S^{\dagger}\left(0,0\right)=1
\label{Sp evolution operator initial condition}\end{equation}
and the group property
\begin{equation}
S\left(t_{1},\bar{t}\right)S\left(\bar{t},t_{2}\right)=S\left(t_{1},t_{2}\right).
\label{Sp evolution operator group property}\end{equation}
Specializing to the case of $G^{>}$, we get:
\begin{eqnarray}
&\hbox{}&i\partial_{t_{1}}G^{>}\left(t_{1},t_{2}\right)=\mathfrak{h}\left(t_{1}\right)S\left(t_{1},0\right)g^{>}\left(t_{1},t_{2}\right)S^{\dagger}\left(t_{2},0\right)\nonumber\\
&\hbox{}&+S\left(t_{1},0\right)i\partial_{t_{1}}g^{>}\left(t_{1},t_{2}\right)S^{\dagger}\left(t_{2},0\right)\nonumber\\
&\hbox{}&=\mathfrak{h}\left(t_{1}\right)G^{>}\left(t_{1},t_{2}\right)+I_{1}^{>}\left(t_{1},t_{2}\right).
\label{KBE greater gauge}\end{eqnarray}
where the second equality comes from the Kadanoff-Baym equation for $G^>$. This
results in
\begin{equation}
i\partial_{t_{1}}g^{>}\left(t_{1},t_{2}\right)=
S^{\dagger}\left(t_{1},0\right)I_{1}^{>}\left(t_{1},t_{2}\right)S\left(t_{2},0\right).
\label{KBE of gauged G}\end{equation}
Therefore, by integrating from present time $T_p$ to $T_p+\Delta$,
\begin{eqnarray}
&\hbox{}&i\left[g^{>}\left(T_p+\Delta,t_{2}\right)-g^{>}\left(T_p,t_{2}\right)\right]\nonumber\\
&\hbox{}&=\int_{T_p}^{T_p+\Delta}S^{\dagger}\left(\overline{t},0\right)I_{1}^{>}\left(\overline{t},t_{2}\right)S\left(t_{2},0\right)d\overline{t}\nonumber\\
&\hbox{}&=S^{\dagger}\left(T_p,0\right)\!\!\int_{0}^{\Delta}\!\!\!\!S^{\dagger}\left(\overline{t}+T_p,T_p\right)I_{1}^{>}\left(\overline{t}+T_p,t_{2}\right)S\left(t_{2},0\right)d\overline{t}.
\nonumber\\
\label{Formal integration of gauged G}\end{eqnarray}
Thus
\begin{eqnarray}
&\hbox{}&G^{>}\left(T_p+\Delta,t_{2}\right)=S\left(T_p+\Delta,T_p\right)G^{>}\left(T_p,t_{2}\right)\nonumber\\
&\hbox{}&\!\!\!\!\!\!\!\!\!-iS\left(T_p+\Delta,T_p\right)\!\!\int_{0}^{\Delta}\!\!\!\!S^{\dagger}\left(\overline{t}+T_p,T_p\right)I_{1}^{>}\left(\overline{t}+T_p,t_{2}\right)\!\!.
\label{Formal integration of G}\end{eqnarray}
Up to this point we have not made any approximations but merely formal
rewritings of the KBE. To calculate the $S\left(T_p+\Delta,T_p\right)$
we divide the interval $\left[T_p,T_p+\Delta\right]$
into $N$ intervals in which the single-particle Hamiltonian, $\mathfrak{h}$, is taken constant and evaluated at the midpoint. For a constant
$\mathfrak{h}$ we obtain
\begin{eqnarray}
&\hbox{}&\!\!\!\!\!\!\!\!\!\!\!\!\!\!S\left(T_p+\frac{\left(j+1\right)\Delta}{N},T_p+\frac{j\Delta}{N}\right)\nonumber\\
&\hbox{}&\qquad\quad=\exp\left(-i\mathfrak{h}\left(T_p+\frac{\left(j+1/2\right)\Delta}{N}\right)\frac{\Delta}{N}\right)\!\!,
\label{Sp evolution operator for constant h}\end{eqnarray}
which is evaluated by diagonalization. The resulting expression becomes
\begin{eqnarray}
&\hbox{}&\!\!\!\!\!\!\!\!\!\!\!\!\!\!\!\!\!\!\!\!S\left(T_p+\Delta,T_p\right)=\!\!\prod_{j=0}^{N-1}\!\!\exp\!\left(\!-i\mathfrak{h}\!\left(\!T_p\!+\!\frac{\left(j+1/2\right)\Delta}{N}\!\right)\!\frac{\Delta}{N}\!\right)\!\!.
\label{Sp evolution operator intervals}\end{eqnarray}
Given that $N$ is taken large enough, the only error of the above
expression comes from the extrapolation/interpolation of $\mathfrak{h}$, which
is small as the density (which enters $\mathfrak{h}$ via the HF term) is a 
continuous and smooth function of time. 
To solve Eq. (\ref{Formal integration of G})
we also need to approximate the integral. This can be done in two
different ways depending on which of the two quantities $I_{1}^{>}\left(\overline{t}+T_p,t_{2}\right)$
or $\tilde{I}_{1}^{>}\left(\overline{t}+T_p,t_{2}\right)=S^{\dagger}\left(\overline{t}+T_p,T_p\right)I_{1}^{>}\left(\overline{t}+T_p,t_{2}\right)$ 
is the most slowly varying function. We have tried both and seen that
the $\tilde{I}_{1}^{>}\left(\overline{t}+T_p,t_{2}\right)$ is the smoothest.
The integral is done numerically, typically with a 2- or 4-point formula.
Similar expressions are used for the other KBE. Special attention
is needed only for the time diagonal. In this case we combine the
two first KBE, Eqs. (\ref{KBE greater real}, \ref{KBE lesser real})
and using the property in Eq. (\ref{Relation collision integrals on the time diagonal}), 
\begin{eqnarray}
&\hbox{}&i\left[\partial_{t_{1}}+\partial_{t_{2}}\right]G^{<}\left(t_{1},t_{2}\right)=\left[\mathfrak{h},G^{<}\left(t_{1},t_{2}\right)\right]\nonumber\\
&\hbox{}&\qquad\qquad\qquad+I_{1}^{<}\left(t_{1},t_{2}\right)-I_{2}^{<}\left(t_{1},t_{2}\right),
\label{KBE on time diagonal}\end{eqnarray}
we then change to the variables $t=\left(t_{1}+t_{2}\right)/2$ and $t'=t_{1}-t_{2}$.
This gives
\begin{eqnarray}
i\partial_{t}G^{<}\!\left(t+t'/2,t-t'/2\right)=\left[\mathfrak{h},G^{<}\left(t+t'/2,t-t'/2\right)\right]\nonumber\\
I_{1}^{<}\left(t+t'/2,t-t'/2\right)-I_{2}^{<}\left(t+t'/2,t-t'/2\right)\!.
\label{KBE on time diagonal relative coordinates}\end{eqnarray}
By performing the same gauge transformation as above and setting $t'=t_{1}-t_{2}=0$
we obtain\cite{Twotime}
\begin{equation}
i\partial_{t}g^{<}\left(t\right)=\tilde{I}_{1}^{<}\left(t\right)-\tilde{I}_{2}^{<}\left(t\right),
\label{KBE on time diagonal of gauged G}\end{equation}
where $\tilde{I}^{<}\left(t\right)=S^{\dagger}\left(t,0\right)I^{<}\left(t,t\right)S\left(t,0\right)$.
Integrating from time $T_p$ to $T_p+\Delta$ we obtain
\begin{equation}
\!g^{<}\left(T_p+\Delta\right)=g^{<}\left(T_p\right)-i\int_{T_p}^{T_p+\Delta}
\left(\tilde{I}_{1}^{<}\left(\bar{t}\right)-\tilde{I}_{2}^{<}\left(\bar{t}\right)\right)d\bar{t},
\label{Formal integration of gauged G on time diagonal}\end{equation}
which leads to 
\begin{eqnarray}
&&G^{<}\left(T_p+\Delta\right)=S\left(T_p+\Delta,T_p\right)G^{<}\left(T_p\right)S^{\dagger}\left(T_p+\Delta,T_p\right)\nonumber\\
&&-iS\!\left(T_p\!+\!\Delta,T_p\right)\!\!\int_{0}^{\Delta}\!\!\!\left(\tilde{I}_{1}^{<}\!\left(\bar{t}+T_p\right)-\tilde{I}_{2}^{<}\!\left(\bar{t}+T_p\right)\right)d\bar{t}\nonumber\\
&&\times S^{\dagger}\left(T_p\!+\!\Delta,T_p\right).
\label{Formal integration of G on time diagonal}\end{eqnarray}
The integral is then evaluated in the same way as discussed above for
the case $t_1\neq t_2$. 
\section{TDDFT Exchange-correlation potential from MBPT \label{sec_TDDFT}} 
Our time-dependent densities from the different Many-Body Approximations
also provide insight for the TDDFT exchange-correlation potentials for strongly correlated systems. 
Given a specific approximation, from the resulting time-dependent density we obtain
the corresponding effective potential $v_{eff}=\Sigma_{H}+w+v_{xc}$, where $\Sigma_{H}$ 
is the Hartree potential and $v_{xc}$ the exchange correlation potential. 
In practice, this is done via a numerical reverse engineering procedure.\cite{VerdozziPRL} 
This algorithm imposes that $\left|n\left(R,t\right)-n^{KS}\left(R,t\right)\right|=0$ at each time-step, 
where $n^{KS}\left(R,t\right)=\sum_{\nu}^{occ} | \psi_{\nu}^{KS}\left(R,t\right)|^2$ is the Kohn-Sham density and $\psi_{\nu}^{KS}$ are the Kohn-Sham orbitals. 
The $n^{KS}$ is found by solving $i\dot{\psi}_{\nu}^{KS}=\left(\hat{t}+v_{eff}\right)\psi_{\nu}^{KS}$ 
where the kinetic energy is given by $\hat{t}=-V\sum_{\left\langle RR'\right\rangle \sigma}a_{R\sigma}^{\dagger}a_{R'\sigma}$.
\begin{figure}[th]
\begin{center}
\includegraphics[width=7.7cm,clip=true]{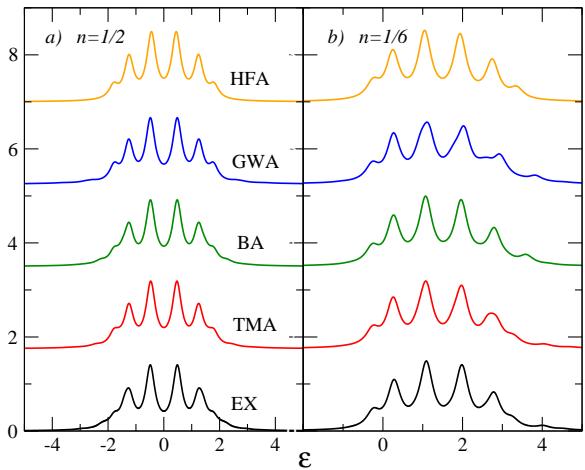}\\
\end{center}
\caption{(Color online) Ground-state spectral functions for $M$=6 for weak interaction strength, $U=1$, for different fillings. The curves correspond to exact (black), TMA (red), BA (green), GWA (blue) and HFA (orange). The curves are shifted for clearer comparison and we have broadened the discrete spectra with a Lorentzian with a $FWHM=0.4$.}
\label{weakU}
\end{figure}
\begin{figure}[t]
\begin{center}
\includegraphics[width=7.7cm,clip=true]{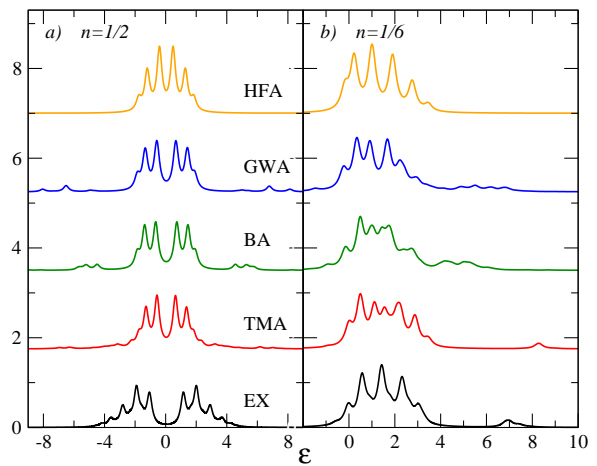}\\
\end{center}
\caption{(Color online) Ground-state spectral functions for $M$=6 for strong interaction strength, $U=4$, at different fillings. The curves correspond to exact (black), TMA (red), BA (green), GWA (blue) and HFA (orange). The curves are shifted for clearer comparison and we have broadened the discrete spectra with a Lorentzian with a $FWHM=0.4$.}
\label{strongU}
\end{figure}
\section{Results: Ground state} \label{sec_results: groundstate} 
To start the time propagation of the KBE, one needs the initial, equilibrium, one-particle propagators. These are found by
by solving the Dyson equation self-consistently. Here, we show only results for isolated clusters,
and defer to a future publication the case of clusters contacted to leads.
To characterize the ground-state properties, we present in Figs. \ref{weakU} and \ref{strongU} 
the spectral functions at the first, $R=1$, site in the cluster. This is the site at which the
external perturbation is applied (see section \ref{sec_results: time dependence}). The results are for the different
Many-Body Approximations, at two particle concentrations and interaction strengths.
A first, general comment is that the performance of MBA:s generally worsens with increasing $U$. 
Some basic, generic features of the different Many-Body Approximations we employ to 
evolve in time our clusters are already revealed by the ground-state results. 
Such features, specifically discussed for the cases in Figs. \ref{weakU} and \ref{strongU}, are 
common to other clusters with different fillings, sizes and interactions 
strengths (such additional results  are not shown).
Fig. \ref{weakU} refers to the situation of weak interaction, where the performance of the MBA:s is satisfactory (in 
particular, all the MBA:s reproduce the symmetry breaking at low filling of the exact solution).
 
At half filling, 
a correlation gap opens on increasing $U$, which becomes clearly visible
in the strong correlation regime, Fig. \ref{strongU}. This behavior in clusters
is consistent with the exact solution of the 1D Hubbard model in the thermodynamical
limit.\cite{LiebWu67} This feature is reproduced  by the different approximate schemes.
However, the size of the approximate gap depends on the Many-Body Approximation:
the best estimate is given by the BA value;  yet, the latter largely
underestimates the exact value. 
Missing the contribution of correlation effects, the non-magnetic HFA solutions reproduce 
the non-interacting spectral density.
As a second general feature, the MBA:s introduce spurious satellites away from the band region.
This problem is most pronounced in the BA and GWA curves.
In the low density regime, for large $U$, a satellite structure appears in the exact solution (about 6.5 in the bottom of panel 4$b$). 
In an extended system, this high energy spectral feature represents a two-electron anti-bound state (i.e. outside the band continuum).  
The satellite is well reproduced by the TMA (although its distance from the band region is overestimated) while is 
smeared out in the BA and GWA and obviously  absent in the HFA (as it includes no correlation effects).
For the strong interaction case, the agreement of the different approximations with the exact curve in the band region is
only moderate.

When comparing the SGWA to the GWA we see a slight improvement, which is expected, 
as the SGWA includes fewer faulty diagrams. This improved version of the GWA is, 
however, still worse than the BA or the TMA. 
In this section, we show no results for the SGWA, since it introduces only a 
a marginal improvement in the ground-state spectral density: however, we will present below SGWA time-dependent densities. 
It is worth noticing that, using a MBA expansion in terms of non-magnetic propagators, 
the SGWA will have a magnetic instability on increasing $U$. This can be seen most easily in a Hubbard dimer with two electrons 
with opposite spins.  In the dimer, where the poles of $W[G_0]$ are $\epsilon=\pm\sqrt{4V^2 \pm 2VU}$, 
this unphysical symmetry breaking occurs for $U \ge 2V$. Conversely, the exact ground state for a dimer is always a spin $S=0$ (singlet) state,
since, for any positive $U$, the dimer ground-state energy
$E^{dimer}_{gs}=E^{S=0}_{dimer}= \frac{U-\sqrt{16V^2+U^2}}{2} < 0= E_{dimer}^{S=1}$.

\begin{figure}[h]
\begin{center}
\includegraphics[width=7.7cm,clip=true]{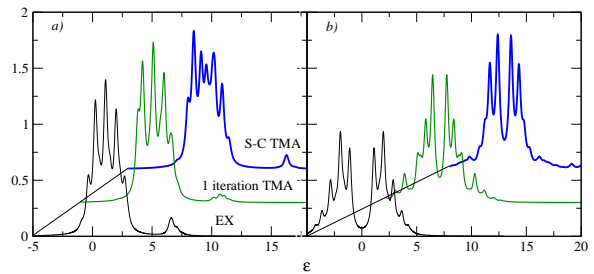}\\
\end{center}
\caption{(Color online) First iteration and self-consistent TMA spectral function versus the
exact one. $M$=6 and in $a)$: $n=1/6$ and in $b)$: $n=1/2$. The curves correspond 
to exact (black), first iteration TMA (green) and self-consistent TMA (thick blue). 
The curves are shifted for clearer comparison and we have broadened the discrete spectra with a Lorentzian with a $FWHM=0.4$.}
\label{1SHvsSC}
\end{figure}

It is worth noting that all the spectral functions shown in Figs. \ref{weakU} and \ref{strongU} 
are not as  good as those obtained without self-consistency, i.e. when stopping after the first iteration of the Dyson equation, see Fig \ref{1SHvsSC}. 
This is an example of the known fact
that self-consistent conserving approximations often have worse spectral properties than 
non self-consistent ones.\cite{Almbladh,Holm} 
To guarantee the fulfillment of the conservation laws and to get unambiguous total energy results it is 
capital to achieve self-consistency. If the quantity of interest is the spectral functions one should instead  
make use of different partial summation criteria,\cite{Almbladh} and in some cases
include vertex corrections to remove artifacts introduced by self-consistency. \cite{MCCV87}

As a final remark to this section we note that all MBA:s based on partial summations involve infinitely 
many possible excitations. These excitations, represented by diagrams in the self energy, result in 
infinitely many, but discrete, number of poles in the ground-state 
spectral functions.\cite{ContinuosA} 
The exact solution, in contrast, lives in a finite phase space which implies a finite number of poles.

\section{Results: Time dependence \label{sec_results: time dependence}}

In this section we examine the performance of the different MBA:s. To accomplish
this, we use as benchmark exact many-body solutions. In general, the latter 
are available only for finite systems, and numerical in nature. As a consequence, 
in this section we deal exclusively with isolated clusters, and focus on general
aspects of the time-dependent densities and MBA:s. However, we defer to the next section
two important outcomes of the KBE time evolution: the correlation-induced damping and
the existence of multiple steady states in isolated and contacted clusters.

{\it Isolated clusters: MBA:s vs exact results.} 
We start the time evolution at $t = 0$ with the ground-state Green's function. For positive times $t > 0$ we 
apply a spin-independent external field to the system. We have studied different types of external fields, but in this paper
we present results only for the form $w_R\left(t\right)=w_{0}\delta_{R,1}\Theta$$\left(t\right)$, that is we consider
a step perturbation and let it to act only on the leftmost, $R=1$, site.
The time is given in units of the inverse hopping parameter $(1/V)$ and all curves 
represent the dynamics on site $R=1$. 
The cases displayed are the same as those considered for the ground-state results.

\begin{figure}[h]
\begin{center}
\includegraphics[width=7.7cm, clip=true]{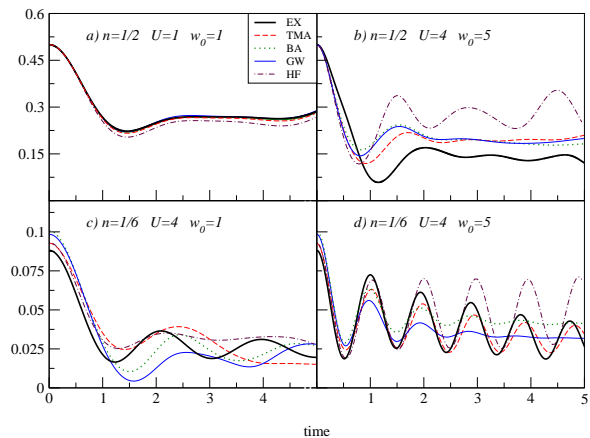}\\
\end{center}
\caption{(Color online) Time-dependent densities on site 1 for $M=6$. The curves correspond to exact (thick solid black), TMA (dashed red), BA (dotted green), GWA (thin solid blue) and HFA (brown dashed dot).}
\label{td}\end{figure}

We show the resulting time-dependent densities in Fig. \ref{td}. 
The curves correspond to some initial states shown in Figs. \ref{weakU} and \ref{strongU}.
In the panels a) and b), $n=1/2$ while in panels c) and d), $n=1/6$. In the simplest case ($U=1, w_0=1$), 
presented in panel a), all MBA:s give a good description of the density.
On increasing the strength of the interaction and the external field, panel b), 
we clearly see that the HFA description is rather crude, 
whilst the curves from the other MBA:s are very similar to each other and closer to the exact density. 
In the case of strong interaction but weak field, panel c), we see that none of the MBA:s 
give an adequate description. We interpret these results as a consequence to the 
fact that the ground-state spectral 
function is not well described in the band region: the latter is responsible for the response to weak fields. 
On the contrary, when the field is strong, panel d), the TMA performs much better than the other MBA:s. 
This is due to the fact that the non-linear response involves states at 
higher excitation energy and the TMA is the only
approximation which, to some extent, reproduces the satellite structure. \newline

\begin{figure}[h]
\begin{center}
\includegraphics[width=7.7cm, clip=true]{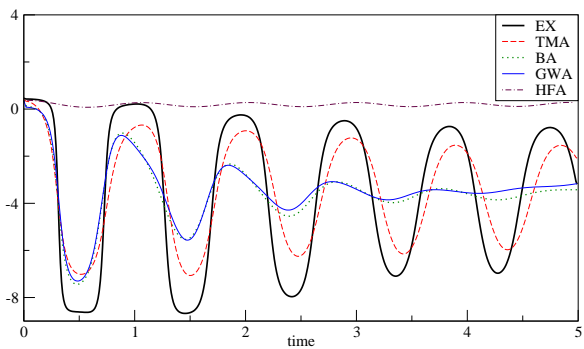}\\
\end{center}
\caption{(Color online) Time-dependent $v_{eff}$ on site 1 for $M=6$,\, $U=4$ and $w_0=5$. The curves correspond to exact (thick solid black), TMA (dashed red), BA (dotted green), GWA (thin solid blue) and HFA (brown dashed dot).}
\label{TDDFT}\end{figure}

{\it Isolated clusters: MBA:s, exact results and TDDFT.} 
It is interesting to examine some of the results just presented from a TDDFT 
perspective. A clear advantage of TDDFT is that, for the time evolution, it deals with
quantities with a single time argument (one is propagating the Kohn-Sham orbitals).
Nevertheless, a key requirement in TDDFT is that $v_{xc}$ (an thus $v_{eff}$) should
depend in a non-local (in space and time) fashion on the particle density.
In this way, all the complexities of the many-body dynamics are subsumed  into
a highly non-trivial dependence of $v_{xc}$ on the density.
Not surprisingly, making progress in the construction of improved XC functional 
is a rather challenging task,
especially for strongly correlated systems. It is also true that, 
in some cases, simple adiabatic local approximations
can provide satisfactory results; this is also the case for a TDDFT description
of the Hubbard model in non equilibrium.\cite{VerdozziPRL, Polini08}
However, memory and non-local effects should in general be taken into account.
The so-called variational approach to TDDFT,\cite{Varia1,Varia2} has the advantage of a systematic
inclusion of many-body contributions in the XC potential. Also, in this way non-locality
in space and memory effects are properly included, once  $v_{xc}$
is retrieved from the many-body self energy via the time-dependent Sham-Schl\"uter equation.\cite{RobertShamSh96}
Deferring a "bottom-up" construction of $v_{xc}$ via MBA:s to future work,  we here still
wish to explore the connection between TDDFT and MBA:s on the Keldysh contour,
looking at exchange-correlation potentials obtained via time-dependent reverse engineering
using the time-dependent densities from  the KBE. The results of this procedure are presented in
in Fig. \ref{TDDFT}, $M=6, N=2, U=4, w_0=5$. This corresponds to the time-dependent densities
presented in panel d) of Fig. \ref{td}. Consistently with the density results, we see
the $v_{eff}$ in the TMA is superior to those from the other MBA:s, and quite close to the exact one.
We also note the large discrepancy of the HFA and that, for both the BA and the GWA,
$v_{eff}$ exhibits a damped behavior (the same is observed in the densities in panel d) of  
Fig. \ref{td}, see section \ref{sec_damping} below).  In spite of not being perfect, the 
agreement of $v_{eff}$ from the TMA with the exact one is quite encouraging, suggesting
that there is ample scope for pursuing the construction of improved $v_{xc}$ from
(suitably chosen) MBA:s.

\begin{figure}[h]
\begin{center}
\includegraphics[width=7.7cm,clip=true]{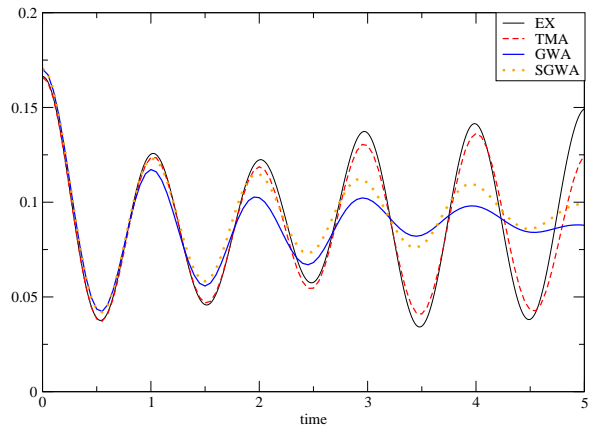}\\
\end{center}
\caption{(Color online) Time-dependent densities for $M=4,\, U=1.5,\, w_0=5$ and $n=1/4$. 
The curves correspond to exact (black), GWA (dotted red), SGWA (thick blue) and TMA (dashed green).}
\label{GWAvsSGWA}\end{figure}

{\it Spin-dependent GWA.} Before concluding this section, we wish to discuss briefly the effect on making a spin-dependent treatment 
in the GWA.  In Fig. \ref{GWAvsSGWA}, we make a comparison between  GWA, SGWA and, for reference,
TMA. Due to the symmetry breaking of the SGWA (see section \ref{sec_results: groundstate}),
and that we only consider the spin unpolarized
case (otherwise, one should start with polarized propagators) we confine ourselves to the weak interaction regime.
As evident from Fig. \ref{GWAvsSGWA}, for the chosen parameters the SGWA is slightly better than 
its spin-independent counterpart but inferior to the TMA.

As a final remark to this section, we note that in the general case, GWA was designed to give a good screening 
of the Coloumb interaction,\cite{Hedin} which in our system has already been taken into account 
indirectly by the model itself. The TMA, on the other hand, is known to give a good performance
if the interaction is short ranged, especially in the low density regime.\cite{TMA}
The general good description of the the TMA (both in and out of equilibrium) for the short-ranged Hubbard
Hamiltonian is in accordance to previous studies of ground state properties of clusters.\cite{MCCV87, CV95}
One can in other words say that the performance of the different approximations 
in the ground state has non negligible relevance to the time-dependent behavior of the system,
and this is especially the case for two most distinctive spectral features, that is:
the band-gap and satellite structure.

\section{Results: Damping and multiple steady states \label{sec_damping}}

\begin{figure}[h]
\begin{center}
\includegraphics[width=7.7cm,clip=true]{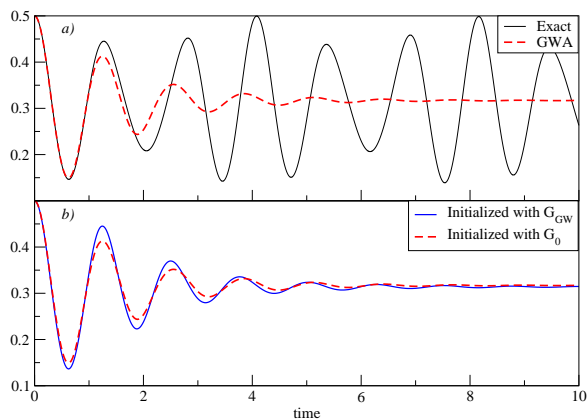}\\
\end{center}
\caption{(Color online) Densities for $M=2,\, n=1/2,\, U=1,\, w_0=5$, exact (black), GWA (dashed red) and SGWA (blue). In $a)$: damping of GWA density versus exact solution. In $b)$: time-dependent densities for the GWA, initialized with the self-consistent GWA ground state and the non-interacting $G_0$.}
\label{damping}
\end{figure}

In this section we will investigate the correlation-induced damping and the multiple solutions 
of the stationary KBE. We find that these features are general for all MBA:s which include correlation 
effects. We exemplify this by presenting results for different MBA:s in the various parts of this section. 

When we let our isolated system(s) evolve under the action of a strong external field 
we reach an artificial steady state,\cite{steadystate, pva} see Fig. \ref{damping}($a$).
The damping mechanism is not a mere consequence of the infinite number of poles in the initial state 
but is rather intrinsically linked to the time propagation scheme. 
To show this fact we present in Fig. \ref{damping}($b$) the time evolved density initiated 
with the non-interacting propagator $G_0$,\cite{initG0} which has a finite number of poles.
>From the curves in Fig. \ref{damping}($b$) it is evident that the non-interacting 
initial state leads to a very similar damped density profile.
Note that the similarity of the curves in Fig. \ref{damping}($b$), indicate a robustness of the KBE time 
evolution against the initial conditions.

\begin{figure}[h]
\begin{center}
\includegraphics[width=7.7cm,clip=true]{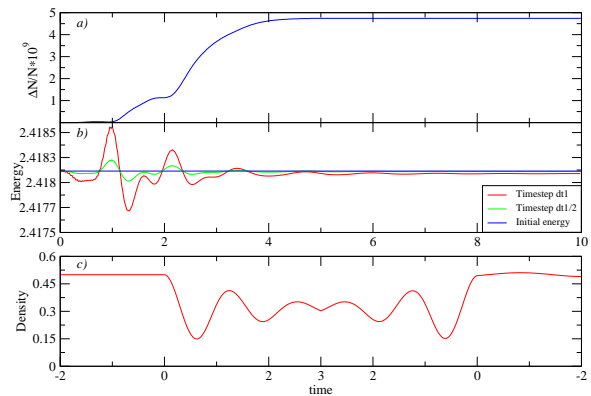}\\
\end{center}
\caption{(Color online) Conservation laws and time reversal in the GWA for 
$M=2,\, n=1/2,\, U=1,\, w_0=5$.}
\label{numerics}
\end{figure}

The damping is not a numerical artifact: in Fig. \ref{numerics} we see that particle and energy conservation 
are strictly obeyed within our numerical accuracy, see top and middle panel. Moreover, the evolution satisfies time-reversal symmetry. 
That is, when we reverse the direction
of time in the propagation, the system goes back to the initial state and 
remains there, see bottom panel. 
The damping rate increases with the strength of the 
external field and is absent in the regime of linear response. The dynamics in this 
limit is described by the
Bethe-Salpeter equation, with a kernel $\delta\Sigma/\delta G$. The
latter would have a discrete spectrum in our MBA:s, and so would the
resulting density response. A discrete response function will in turn lead to a non-damped dynamics. 
We wish to stress that in a formulation based on  conserving Many-Body Approximations, the key quantity is 
the generating functional; the corresponding $G$ is 
then defined via the self-consistent KBE, and not via an underlying wave function scheme. Without the 
connection to the underlying wave function, there is no guarantee 
that, for example, systems with a finite phase space will only have a finite number of excited states
or that they will not have a damped dynamics.

\begin{figure}[h]
\begin{center}
\includegraphics[width=7.7cm,clip=true]{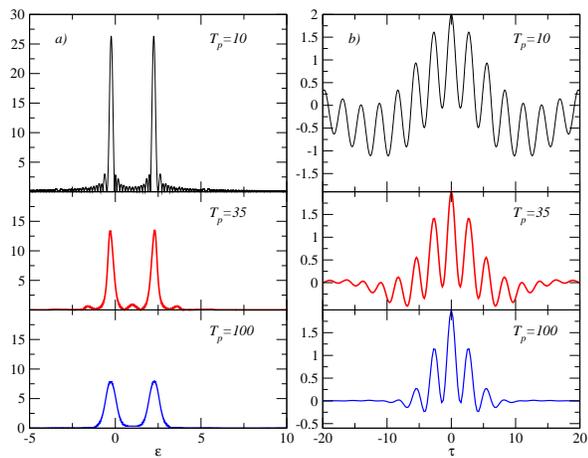}\\
\end{center}
\caption{(Color online) Time-dependent spectral functions in the GWA for $M=2,\, n=1/2,\, U=1,\, w_0=2$, $a)$: in energy space and $b)$: in time space. The different curves correspond to $T_p=10$ (black), $T_p=35$ (red) and $T_p=100$ (blue).}
\label{tdspectrum}
\end{figure}

To study the non-linear response regime, we find convenient to introduce the instantaneous spectral function:  
\begin{equation}
A\left(T_p,\omega\right)\!=\!-\textup{Tr}\: \textup{Im}\!\!\int_{-2T_p}^{2T_p} 
\!\!\!\!\!\!e^{i\omega \tau}\left[G^{>}\!\!-\!G^{<}\right]\!\left(T_p\!\!+\!\!\frac{\tau}{2},T_p\!\!-
\!\!\frac{\tau}{2}\right)\!d\tau ,
\end{equation}
where $T_p=\left(t_{1}+t_{2}\right)/2$ and $\tau=\left(t_{1}-t_{2}\right)$, 
and its counterpart in time space.
When reaching the steady state, the spectral function
gets broadened in energy space, Fig. \ref{tdspectrum}($a$), and damped in time space, 
Fig. \ref{tdspectrum}($b$). Note that, for our dimer, the exact instantaneous spectral 
function would continue to oscillate in time space. 
The different MBA:s have different damping rates; among them, the TMA is in general the slowest.
The damping acts strongest on the perturbed site, it generally
decreases with system size and is most important at half filling.
In our study we have not made an exhaustive characterization of how the large-size limit is gradually obtained;
this is indeed an the issue we plan to address in future work. Still, we wish to present in the rest of this section
some general considerations on the effect of system-size on the damping behavior.\newline
{\it System size and damping}. In an exact treatment, a finite system has a corresponding finite phase space and will thus not be able to 
fully relax to a stationary steady state. In a large but finite phase space it will, due to decoherence, give rise to a quasi steady state.
The larger the system gets, the more and more complete the damping becomes. 
Thus, after a long time, in an exact time dynamics the system would 
exhibit noise-like fluctuations which never die but which decrease in amplitude
with system size. We have seen that in our approximate KBE evolution
the system, while being finite, attains a stationary state.
Thus, the artificial damping of the KBE dynamics is expected 
to increasingly resemble the physical damping as the system grows. \newline
{\it Self-consistency and damping}. In Many-Body Perturbation Theory, the self energy accounts for possible excitations 
of the system which involve a certain number of particles/holes. 
Any MBA based upon partial summations includes diagrams of all orders. 
These terms act as an effective bath which gives rise to infinitely many 
discrete poles in the ground state (in isolated systems) and correlation-induced damping in the time dynamics.
In a finite system, there will be contributions to the self energy which annihilate more holes/particles than 
those which can be accommodated in the system. 
In an exact theory, these unphysical terms would be exactly canceled, order by order, by other unphysical 
pieces. In approximations such as the GWA or the TMA, there is in general 
no such perfect compensation. The infinite number of poles of the ground state and 
the correlation-induced damping will thus be artificial for a finite system. 

\begin{figure}[h]
\begin{center}
\includegraphics[width=7.7cm,clip=true]{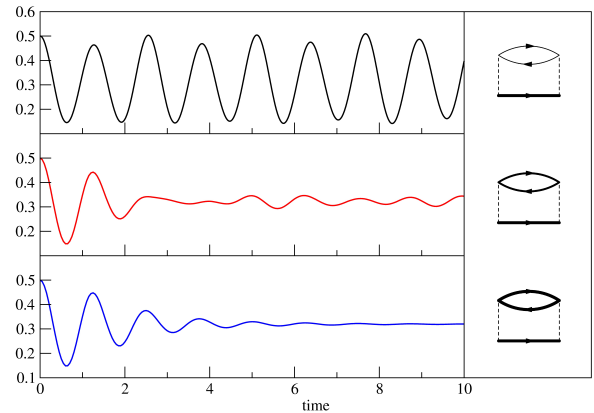}\\
\end{center}
\caption{(Color online) Time-dependent densities for Born approximations with different levels of self-consistency 
for $M=2,\, n=1/2,\, U=1,\, w_0=5$; BA$_0$ (black), BA$_{\rm{HFA}}$ (red) and BA (blue).}
\label{scBA}\end{figure}
In Fig. \ref{scBA}, we present results from three versions 
of the BA to illustrate the effect of an increasing level of self-consistency. 
These approximations, which are all particle-conserving,\cite{Ulf} involve different polarization propagators.
In the first case we evaluate the polarization propagator with ground-state propagators 
(BA$_0$) (top panel). In this case the density does not damp.
In the second case we evaluate the polarization with propagators in the 
time-dependent HFA approximation (BA$_{\rm{HFA}}$) (middle panel). In this case we observe partial damping.
If we finally use the self-consistent $G$ (BA) (bottom panel), we get complete damping. 
We see in other words that if all $G$:s
that build up the the self energy are the self-consistent ones we reach a steady state.\cite{except} \newline
{\it Multiple steady states.}
\begin{figure}[h]
\begin{center}
\includegraphics[width=7.7cm,clip=true]{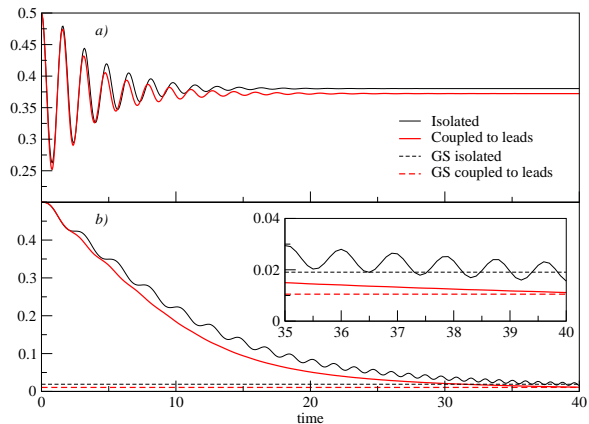}\\
\end{center}
\caption{(Color online) Time-dependent densities for a cluster isolated and coupled to leads in the TMA 
with sudden and slow switch-on. $M=2,\, U=1,\, w_0=8$. 
Lead parameters: $V_{L}$ = 1, $V_{LC}$ = 0.5, filling = 0.5 and bias = 0. $a)$ Sudden 
switch-on: isolated cluster (thin black), coupled to leads (thick red). $b)$ Slow 
switch-on: isolated cluster (thin black), coupled to leads (thick red); ground-state 
densities of final Hamiltonian: isolated cluster (dashed black), coupled to 
leads (dashed thick red). {\it Inset}: expanded view of the long time limit behavior.}
\label{Multss}\end{figure}
Another striking feature related to the correlation-induced damping is that the steady state is not unique for a given final 
external field: it depends on how the perturbation is switched on. 
In our simulations, in the case of adiabatic turn-on, we reach the ground 
state of a system with an on-site energy corresponding to the final external perturbation. 
This is consistent with the adiabatic theorem. If, however, the perturbation is switched on suddenly, 
we reach a non-physical steady state with the same energy as at $t=0^+$. 
This non uniqueness is indicative of an important aspect: given a final external potential, there are
multiple, in principle infinitely many, solutions of the stationary KBE in finite clusters.
Furthermore, as we show next, this general statement, in some cases, remains valid
in presence of macroscopic leads, where we have both a self energy $\Sigma_{emb}$ from
the leads and a self energy $\Sigma_{MBA}$ from the interactions. Both contributions
are non-local in time and may lead to damping. 
Thus, in general we have two damping mechanisms: One due to the coupling to the 
continuous lead band and one induced by correlations.

In Fig. \ref{Multss} we present the time-dependent densities within the TMA for a dimer, both 
isolated and coupled to unbiased leads,\cite{Leads,Hopping} subject to an external field with sudden and 
slow ($w\left(t\right)=w_{0}t/t_{max}$) switch-on, (panel $a)$ and panel $b)$ respectively). 
 
The strength of the perturbation is such that the ground state corresponding 
to the final external potential contains states outside the band continuum (bound states);
in this case we find that the KBE admit multiple steady-state solutions.\cite{bias} 
These steady states vary continuosly with the way the final potential is reached. 
In other words we have infinitely many steady states where similiar switch-ons give 
similar steady states. In the case of the HFA we see that the density continues to 
oscillate if we have more than one pole in the ground state of the final Hamiltonian.
This results is consistent with recent work on the role of bound states in quantum transport.
\cite{bound1,bound2,bound3,bound4} However, we wish to remark that, in the HFA,
the frequency, amplitude and average value of the oscillating 
density will, however, depend on the way the external field is switched on. \cite{Frequency} 

Similarly to the isolated case, we find
that for a slow switch-on we tend to reach the ground state corresponding to the final 
Hamiltonian (given by the dashed curves), while for the sudden switch-on we reach another steady state.
If the external field is such that the corresponding ground state of the final Hamiltonian 
includes only excitations within the band then the steady state reached is the final ground 
state, independently on how the perturbation is switched on.   

\begin{figure}[h]
\begin{center}
\includegraphics[width=7.7cm,clip=true]{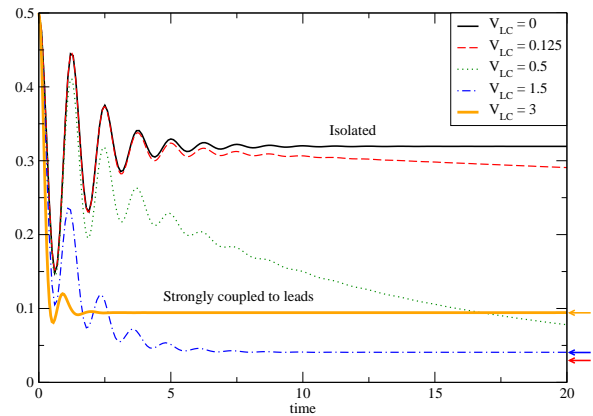}\\
\end{center}
\caption{(Color online) Time-dependent densities for a cluster with different 
couplings to unbiased leads in the BA. $M=2$, $U=1$, $w_0=5$, $V_{L}=5$.}
\label{vlink}\end{figure}
The damping induced by correlations and the one due to the coupling to the 
macroscopic contacts have in general different characteristic time scales. This is clearly 
seen in Fig. \ref{vlink} where we have gradually increased the coupling to the leads: here,
the  external field and the lead band width are such that the corresponding ground state of the 
final Hamiltonian contains no poles. In the case of an isolated cluster ($V_{LC}=0$), 
the damping is only due to correlations. Once the coupling to the leads is non-zero, 
there will be, in addition to the correlation-induced one, a damping due to the 
contacts; the latter will eventually bring the system to the corresponding ground state. 
In Fig. \ref{vlink} , the ground-state values of the densities are represented by arrows
(on the scale of the figure, the red and green arrows are not distinguishable). Note that
each curve in Fig. \ref{vlink} corresponds to a different system, i.e. with a different device-lead 
coupling strength. Each of these systems has a different final Hamiltonian, i.e. a different 
corresponding ground state.
When the coupling strength is weak ($V_{LC}=0.125, 0.5$), the time scale of the damping due to the leads 
is much longer than correlation-induced one. For intermediate coupling strengths ($V_{LC}=1.5$), 
the characteristic times will be of the same order of magnitude and the interplay of the two mechanisms 
becomes intrisically difficult to discern. When the coupling strength is strong ($V_{LC}=3$), 
the damping is completely dominated by the leads.

As a final remark, we know that the correlation-induced damping is 
artificial in isolated clusters, and we saw in Fig. \ref{vlink} that when the coupling to the leads 
is weak, the initial damping is also dominated by correlations. These two facts together 
seem to cast some doubt on the capability of the KBE+MBA:s  scheme to describe
(within the simple MBA:s  discusses here) transients in the weak coupling case 
(a regime which is of high experimental interest, and in fact  among the most investigated in the literature).

To summarize this section, we see that correlation-induced damping and multiple 
steady states are present both in isolated and contacted clusters. 
In the case of isolated clusters, this damping is artificial as the 
exact solution does not reach a steady state. 
The question whether or not the 
correlation-induced damping and the existence of multiple steady states in a cluster coupled to 
leads is an artifact of Many-Body Perturbation Theory is beyond the scope of
this paper and is left to future work.     
\section{Conclusions and outlook \label{sec_conclusions}} 
A main objective of this paper has been to describe in detail a method, within 
the framework of the time-dependent Kadanoff-Baym equations (KBE),
to study finite systems in equilibrium as well as out of equilibrium. The main
emphasis of the paper has been on finite clusters, for which
a meromorphic representation of the equilibrium many-body quantities is possible, but, 
in few instances, we have also considered some results for contacted clusters.

As a concrete application, we examined the time evolution of clusters with strong, short ranged 
electron interactions within several Many-Body Approximations, and compared the results to 
exact ones. A first main outcome of this comparisons is that, for short ranged,  
Hubbard-type interactions, the T-matrix approximation performs very well at low densities, 
and is in general superior to the GW and second Born approximations, both in describing 
the time-dependent density and the corresponding exchange-correlation potential.

A second important outcome of our work is the existence
of two remarkable  features of the time-dependent KBE. The first is that the KBE 
present a correlation-induced damping in the non-linear regime. The second is that the 
steady state reached is in general not unique, i.e. the stationary KBE support multiple stationary states.
Since a finite cluster subject to a non-adiabatic perturbation will oscillate indefinitely, it is clear 
that, for finite systems, such damping and multiple steady-state behavior are artificial.
We argue that this shortcomings will always be present when applying infinite order perturbation theory based on partial
summations to finite systems. In this paper, we were not able to provide a conclusive answer to
whether such two aspects of the KBE bear any physical meaning or if they are a spurious effect of MBPT.
As we discuss next, this is part of our plans for follow-up work.

Future research activity may be envisaged in various directions.
On the methodological side, it would be of interest to further
investigate, for contacted systems, 
how sound the correlation-induced damping and the multiple steady 
states are. \cite{possiblestrategy}

A second methodological issue we are currently addressing is  
performance of the different MBA:s in quantum transport geometries. 
We will will do by comparing the approximate results with those of time-dependent 
density-matrix renormalization group calculations.\cite{Schmitteckert}

Another possible line of study would be searching for algorithms to reduce/optimize the 
computational costs of time-dependent KBE numerical calculations. This is necessary 
to deal with more realistic systems. 

As a fourth  direction, it would be of great interest to study the effect of including 
bosonic degrees of freedom as vibration phenomena often play an crucial role in quantum transport. 

Finally, we also intend to use our KBE-based computational treatments in some specific applications: Possible
examples are real-time qubit manipulation, bistability induced by electron-phonon interaction, cold-atom dynamics.
More specific details are deferred to future publications.

\section*{Acknowledgments \label{sec_acknowledgements}}
Fruitful discussions with Ulf von Barth, Gianluca 
Stefanucci and Robert van Leeuwen  are gratefully acknowledged. This work was supported by the 
EU 6th framework Network of Excellence NANOQUANTA (NMP4-CT-2004-500198) and the 
European Theoretical Spectroscopy Facility (INFRA-2007-211956).

\appendix
\renewcommand{\theequation}{A-\arabic{equation}}
\setcounter{equation}{0}  
\section*{APPENDIX A: Energy functionals \label{sec_appendix}}
The energy functionals considered here are the Galitskii-Migdal (GM) and the
Luttinger-Ward (LW). 
An important difference between them
is that the GM needs only the knowledge of the single-particle Hamiltonian
and of the spectral function of $G$, while the LW incorporates the $\Phi$
functional as well as $\Sigma\left[G\right]$, which depends on
which approximation one is using. Here we give the expression
for the LW functional for the GW-approximation. In practice, we evaluated the LW only
in the ground state, i.e. in frequency space, while the GM was also used during
time evolution.  
\subsubsection{Galitskii-Migdal}
In frequency space, the GM functional reads \cite{Galitskii-Migdal}
\begin{equation}
E=-\frac{i}{2}\, \textup{Tr}\left\{ \left(\epsilon+h\right)G\right\}, 
\label{GM functional}\end{equation}
where $\textup{Tr}$ stands for trace \cite{trace} 
and $h$ is the single-particle Hamiltonian {\it not} including $\Sigma_{HF}$. Thus
\begin{equation}
E=\sum_{R}\sum_{i=1}^{n_{F}}a_{i}\, A_{RR}^{i}-\sum_{RR'}\sum_{i=1}^{n_{F}}A_{RR'}^{i}\, h_{RR'},
\label{GM expression}\end{equation}
where $n_{F}$ is the number of poles below the Fermi energy, and after having
made use of the meromorphic representation of $G^<$. 
In time space, the form of E becomes
\begin{equation}
E=-\frac{i}{2}\textup{Tr}\left[\left(i\partial_{t}+h\right)G^{<}\left(t,t^{+}\right)\right].
\label{GMtime}\end{equation}
Making use of  the the equation of for $G$, we obtain
\begin{eqnarray}
&\hbox{}&E=-\frac{i}{2}\textup{Tr}\left[hG^{<}\left(t,t^{+}\right)\right]\nonumber\\
&\hbox{}&-\frac{i}{2}\textup{Tr}\left[\left(h+\Sigma_{HF}\right)G^{<}\left(t,t^{+}\right)+I_{1}^{<}\left(t,t^{+}\right)\right].
\end{eqnarray}

\subsubsection{Luttinger-Ward}

The LW energy functional is \cite{Luttinger-Ward}
\begin{equation}
iE=\Phi-\textup{Tr}\left\{ \Sigma G+\textup{ln}\left(\Sigma-G_{0}^{-1}\right)\right\}, 
\label{LW functional}\end{equation}
where the self energy depends functionally on the input $G$ i.e. $\Sigma=\Sigma\left[G\right]$
and $\Phi$ is the generating (of the MBA:s) functional.
Specializing to the case of the GW approximation, the $\Phi$ functional becomes
\begin{equation}
\Phi_{GW}=\Phi_{HF}+\frac{1}{4}\textup{Tr}\left\{ \textup{ln}\left[1-UP\right]+UP\right\}, 
\label{Generating functional GW}\end{equation}
where $\Phi_{HF}=(i/2)\,\textup{Tr}\,\{ \Sigma_{HF}[G]\, G\}$.
When computing the logarithmic term $\textup{ln}\left(\Sigma-G_{0}^{-1}\right)$ in Eq. (\ref{LW functional}),
it is useful to make the following separation
\begin{equation}
\ln \left[\Sigma-G_{0}^{-1}\right]=\textup{ln}\left[-G_{HF}^{-1}\right]+\textup{ln}\left[1-G_{HF}\Sigma_{c}\right],
\label{Logarithmic term}
\end{equation}
where $G^{-1}_{HF}=\epsilon-\mathfrak{h}$ and $\mathfrak{h}$ incorporates the the 
Hartree and Fock terms and where $\Sigma_c$ is the correlation part of the GW self energy.
The first term in Eq. (\ref{Logarithmic term}) is the sum of the eigenvalues of the occupied Hartree-Fock
single-particle states (calculated from the correlated Green's function)
which obviously does not correspond to the Hartree-Fock energy.
The second contribution in Eq. (\ref{Logarithmic term}) can be evaluated with a well known identity \cite{Trick}: 
\begin{eqnarray}
&\hbox{}&\textup{ln}\left[1-G_{HF}\Sigma_{c}\right]=-\int_{0}^{1}\frac{G_{HF}\Sigma_{c}}{1-\lambda G_{HF}\Sigma_{c}}d\lambda\nonumber\\
&\hbox{}&\qquad\qquad\qquad\qquad=-\int_{0}^{1}\widetilde{G}\Sigma_{c}d\lambda.
\label{Feynman trick}\end{eqnarray} 
In Eq. (\ref{Feynman trick}), 
$\widetilde{G}$ satisfies the Dyson equation
\begin{equation}
\widetilde{G}=G_{HF}+\lambda G_{HF}\Sigma_{c}\widetilde{G}.
\label{Effective Dyson G}\end{equation}
The other logarithmic
term in $\Phi_{GW}$, i. e. $\textup{ln}\left[1-UP\right]$, is treated in a similar way.
In the actual calculation, expressions of the form $\textup{Tr}\left[AB\right]$ are evaluated analytically,
whilst integrals of the kind $\int_{0}^{1}ABd\lambda$  are performed numerically.


\begin{thebibliography}{10}
\bibitem{pva1} A short account of this work was reported previously. \cite{pva}
\bibitem{Hartree} 
In extended Coulomb systems, $w$ and the Hartree potential $\Sigma_{H}$ are infinite 
but with a finite sum and must be treated together. In this work we consider model systems 
with short-range interactions and such a procedure is not mandatory.
\bibitem{electronhole} There exist also contributions coming from ladder diagrams
with interactions only/also in the electron-hole channel, usually of principal relevance
in exciton-type problems; however, such contributions are not the dominant ones in
the low-density limit for our Hubbard clusters with one orbital/site; and in the rest of the paper, we do 
not consider them altogether, limiting ourselves to particle-particle or hole-hole 
ladder contributions.
\bibitem{Dyson} The Dyson equations for $T$ or $W$ are solved in a similar way.
\bibitem{Symmetry} In the case of complex time arguments we would 
have $G\left(12\right)=-G\left(2^{*}1^{*}\right)^{\dagger}$.
This will henceforth be implicit in all similar symmetry relations.
\bibitem{Tsymmetry}The $T^{>}\left(t,t\right)$ is also needed as there is no closed
relation to $T^{<}\left(t,t\right)$.
\bibitem{Realdensity} In general it would correspond to the 
occupation numbers of the single-particle basis.
\bibitem{Twotime} Explicit reference to the two-time dependencies is omitted as $t_{1}=t_{2}$.
\bibitem{ContinuosA} Continuous broadened ground-state spectral functions have 
been found for few electron quantum dots \cite{Bonitzdots}, in contrast, for the 
systems considered here our results show only discrete spectral features.
\bibitem{steadystate} In a steady state all the single-particle
properties are independent of time and the one-particle Green's function
depends only on the time difference, i.e. $G(t,t') \rightarrow G(t-t')$.
\bibitem{except} This conclusion applies to all MBA:s here, although
partially self-consistent, but particle-conserving schemes can be
easily devised for GWA but not for TMA.
\bibitem{Leads} The theoretical formulation to treat macroscopic leads within the Keldysh formalism will be presented in detail in a future paper.
\bibitem{Hopping} The hopping parameter in the central region is, similarly to the isolated case, set to unity throughout the paper.
\bibitem{bias} The existence of multiple steady states remains valid in the case that we apply a bias to the leads.
\bibitem{Frequency} In the case of $U=0$ the frequency is independent of the switch-on.\cite{bound2}
\bibitem{possiblestrategy} 
Should these features of the KBE turn out to be artificial, one could try to find
ways to suppress the non-physical behavior while keeping the connection to MBPT.
One possibility could be to use the variational properties 
of the underlying $\Phi$ functional and use the Sham-Schl\"uter equation to determine on the fly
the corresponding TDDFT potentials. At full self-consistency, the TDDFT 
dynamics would just reproduce the damped KBE dynamics.
If one instead use optimized one-particle $G$ when constructing the exchange correlation potential, 
$v_{xc}$.\cite{Ulfetal,Varia1}, we expect that the non-physical damping will be reduced or removed 
without seriously deteriorating the short-time dynamics. In the BA, this approach consists in
using a (time-) optimized 
independent-particle $G$ when constructing the self energy and $v_{xc}$ and in 
section \ref{sec_damping} 
we saw that the damping is indeed reduced. 
The use of MBPT based TDDFT potentials could be of great importance of its own due to the possibility of treating
larger and more complex systems.
\bibitem{trace} In frequency space  $\textup{Tr} \equiv 2  \lim_{\eta \rightarrow 0^+} 
\sum_{RR'} \delta_{RR'} \int\frac{d\epsilon}{2\pi}e^{i\epsilon\eta}$ and 
in time space $\textup{Tr} \equiv 2  \sum_{RR'} \delta_{RR'}$. 
The space in which the expressions are represented are given by the context.
\end{thebibliography}
\end{document}